\documentclass[reprint,amsmath,amssymb,aps,pra]{revtex4-2} 

\usepackage{graphicx}
\usepackage{epsfig}
\usepackage{epstopdf}
\usepackage{subfigure,color}
\usepackage{dcolumn}
\usepackage{bm}
\usepackage{lineno}

\def\wM{\omega_{\rm M}}
\def\ws{\omega_{\rm s}}
\def\wn{\omega_{n}}

\def\wmd{\omega_{\rm d}}
\def\fs{f_{\rm s}}
\def\fd{f_{\rm d}}
\def\Rhload{\rho_{\rm load}}
\def\Rhsource{\rho_{\rm source}}
\def\Rhpump{\rho_{\rm pump}}
\def\thetad{\theta_{\rm d}}
\def\ZL{\bar{\bar{Z}}_{\rm L}}
\def\YC{\bar{\bar{Y}}_{\rm C}}
\def\ZR{\bar{\bar{Z}}_{\rm R}}
\def\ZM{\bar{\bar{Z}}_{ {M}}}
\def\Rload{R_{\rm load}}
\def\Pload{P_{\rm load}}
\def\Psource{P_{\rm source}}
\def\Ppump{P_{\rm pump}}
\def\Zin{Z_{\rm in}}
\def\mM{m_{\rm M}}

\def\mM{m_{\rm M}}

\def\phiM{\phi_{\rm{M}}}
\def\phiM{\phi_{\rm{M}}}

\def\Wmat{\bar{\bar{W}}}

\def\iLt{I_{\rm L}(t)}
\def\vLt{V_{\rm L}(t)}
\def\Vst{V_{\rm s}(t)}
\def\Zs{Z_{\rm s}}
\def\ZLeff{Z_{\rm L-eff}^{\omega_s}}
\def\RLeff{R_{\rm L-eff}^{\omega_s}}
\def\LLeff{L_{\rm L-eff}^{\omega_s}}
\def\Re{{\rm{Re}}}
\def\Im{{\rm{Im}}}
\def\Pload{P_{\rm load}}
\def\PRt{P_{\rm R1}}
\def\PRr{P_{\rm R2}}
\def\Ps{P_{\rm s}}

\begin{document}
\preprint{APS/123-QED}

\title{Time-varying components for enhancing wireless transfer of power and information}

\author{Prasad~Jayathurathnage$^{1}$}
\author{Fu~Liu$^{1,2}$}
	\email{fu.liu@xjtu.edu.cn}
\author{Mohammad~S.~Mirmoosa$^{1,3}$} 
\author{Xuchen~Wang$^{1}$}
\author{Romain~Fleury$^3$}
\author{Sergei~A.~Tretyakov$^{1}$}

\affiliation{$^1$Department of Electronics and Nanoengineering, Aalto University, P.O.~Box 15500, FI-00076 Aalto, Finland\\
$^2$Key Laboratory for Physical Electronics and Devices of the Ministry of Education and Shaanxi Key Lab of\\
Information Photonic Technique, School of Electronic Science and Engineering, \\
Faculty of Electronic and Information Engineering, Xian Jiaotong University, Xi'an 710049, China\\
$^3$Laboratory of Wave Engineering, Swiss Federal Institute of Technology in Lausanne (EPFL), CH-1015 Lausanne, Switzerland}

\begin{abstract}
Temporal modulation of  components of electromagnetic systems provides an exceptional opportunity to engineer the response of those systems in a desired fashion, both in the time and frequency domains. For engineering time-modulated systems, one needs to thoroughly study the basic concepts and understand the salient characteristics of temporal modulation. In this paper, we carefully study physical models of basic bulk circuit elements -- capacitors, inductors, and resistors --  as frequency dispersive and time-varying components and study their effects  in the case of periodical time modulations. We develop a solid theory for understanding these elements, and apply it to two important applications: wireless power transfer and antennas. For the first application, we show that by periodically modulating the mutual inductance between the transmitter and receiver, the fundamental limits of classical wireless power transfer systems can be overcome. Regarding the second application, we consider a time-varying source for electrically small dipole antennas and show how time modulation can  enhance the antenna performance. The developed theory of electromagnetic systems engineered by temporal modulation is applicable from radio frequencies to optical wavelengths. 
\end{abstract}

\maketitle


\section{Introduction}

Although the concept of ``temporal modulation"~\cite{Faraday} of radio and optical components was propounded decades ago (see, e.g., Refs.~\cite{Cullen,Tien,Morgenthaler,Kamal,Currie,Simon,Edmondson,Oliner,Anderson,Holberg,Tucker}), a great deal of renewed attention has been recently given to this concept, and, today, it is one of the main research topics in electromagnetics community. This is due to the fact that time modulation of (effective) parameters of electric circuits, transmission lines, 
media~\cite{fleurytvmpar,Halevieutv2009,Halevieutv2016,Engheta2020KKR,TTKSFleury2020IEEE}, meta-atoms~\cite{TM-Radiation1}, metasurfaces, metamaterials~\cite{Enghetanano,CalozTVMM1,CalozTVMM2}, etc., can result in overcoming fundamental limitations such as reciprocity~\cite{TM-nonreciprocity1,TM-nonreciprocity2,TM-nonreciprocity3,TM-nonreciprocity4,TM-isolation1, TM-isolation2,FANTretyakov2020,TretyakovAlu2020}, energy accumulation~\cite{TM-Energy}, bandwidth~\cite{TM-BandWidth}, and impedance matching~\cite{TM-Matching}. Moreover, additional functionalities become available, such as frequency conversion and generation of higher-order frequency harmonics of waves~\cite{TM-frequency1}, wavefront engineering~\cite{TM-frequency1,TM-wavefront1,TM-wavefront2}, one-way beam splitting~\cite{TM-beam}, parametric amplification of waves~\cite{Fleury222,Fleury333}, and controlling instantaneous radiation from small particles~\cite{TM-Radiation1,TM-Radiation2}. 

Regarding linear and temporally modulated electric circuits, the reactive elements (capacitors and inductors), and the dissipative element (resistors) are forced to change in time in a desired way. For example, this is the keystone of classical parametric amplifiers in which by periodically modulating a reactive element it becomes possible to amplify signals passing through this element.  In fact, the reason is that such temporal modulation emulates a negative resistance, which allows to partially compensate the internal resistance of the source as well as other losses. Therefore, there is theoretically no fundamental limit on the amplitude of the current drawn from a voltage source (of course, there are practical limitations due to potential instabilities, available power, and so forth). Apart from these parametric systems that  emulate negative resistance, it appears that there are other  potentially useful features and effects in time-modulated electric circuits that have not yet been considered. To realize those characteristics and effects, one needs to establish a general theory and practical design tools that  can be used for thoroughly describing various time-varying elements including dissipative and reactive ones. This paper focuses on this important issue. We introduce a general analytical approach for analysis of time-modulated structures and discuss several intriguing properties that arise due to  temporal variations of circuit elements. We believe that this study  will pave the road toward finding new possibilities or engineering  electromagnetic systems that can effectively and in novel ways use modulated electric circuits for improving their efficiency and performance.

In this paper, we contemplate what a time-varying reactive element means in general, from the  fundamental point of view. We write a linear, causal, and nonstationary relation between the electric charge and voltage, and, consequently, introduce the capacitance kernel. An analogous  relation relates magnetic flux and electric current in terms of the inductance kernel. Based on the notion of these kernels, we define the concepts of ``temporal capacitance" and ``temporal inductance" as well as equivalent reactances. These notions allow us to discuss the fundamental assumptions behind the circuit models of time-modulated components. 
We present a general analytical method for the analysis of linear time-modulated circuits with arbitrary periodic modulation of elements including the resistive element. This method is based on the use of effective matrix circuit parameters (matrix impedances or admittances) which relate frequency harmonics of voltages and currents. 

We employ this approach to study two different possibilities offered by time modulation. As the first possibility, we consider near-field wireless power transfer (WPT) systems in which the temporal modulation of mutual inductance of the two coils allows enhancement of delivered power without compromising power efficiency, which is not possible in conventional systems.
As the second possibility, we find that a time-modulated resistance dramatically affects the total equivalent reactance of an RLC resonant circuit without adding external reactive elements to the circuit. This has a significant impact, for example, in small antennas for tuning and enhancing the antenna bandwidth. Note that such resistance can be provided externally (connected to the input terminals of the small antenna as the internal resistance of the signal source) or by modulating the shape of the antenna.    

The paper is organized as follows. In Section~\ref{sec:basicconcrele}, we give a general explanation of the nature and models of  nonstationary reactive elements, and, in Sections~\ref{sec:gttmeppp} and \ref{sec:tmiprlc}, we discuss the general approach to model and understand the periodically modulated circuit components. In Section~\ref{sec:wpttmmi}, the enhancement of near-field WPT is studied, and, in Section~\ref{sec:tmrapplic}, we show how a time-varying resistor helps to improve  impedance matching. Finally, Section~\ref{sec:conclusion} concludes the paper.   



\section{Basic concepts: time-varying reactive elements} 
\label{sec:basicconcrele}

In this section, we  discuss the assumptions made in modeling bulk (electrically small) components whose parameters vary in time due to modulation by external forces or fields. This discussion is important to define the applicability region of the theoretical models that we develop and use in this paper. Furthermore, we introduce effective parameters of time-modulated elements (temporal capacitance, inductance, and resistance) that will be used in studying enhanced wireless power transfer devices and antenna systems.

\subsection{Time-varying capacitance}

Let us consider a simple  capacitor formed by two metal plates (made of perfect electric  conductors) connected to a time-varying voltage source, which exerts a time-varying electric field $\mathbf{E}(t)$ between them. 
The space between the plates is filled by a dielectric material.  
At this point, we consider an arbitrary temporal function for the voltage source. 

The electric properties of  capacitors can be modulated in time using several means. One approach is to change the geometrical parameters (such as the distance between the plates) by applying some mechanical force. This is possible even if there is no material filling the capacitor. Another approach is to change the properties of the medium that fills the capacitor volume by applying a strong enough external voltage, or illuminating by light, or heating and cooling the capacitor (notice that the medium response to the signal voltage source is still linear). This is the case of a varactor, for example. 

The first assumption that we make is that the capacitor is a bulk element. That is, its size is small compared to the wavelength at all relevant frequencies. This is the usual assumption in the circuit theory, justifying the use of the notions of voltage, magnetic flux, and allowing writing relations between voltages and currents. In the case of time modulations or presence of nonlinear elements, higher-order frequency harmonics are generated, and the size of the component is assumed to be small at all frequencies where the oscillations are significant. Because of the energy conservation and the finite input energy from the modulating mechanism, the amplitude of response at high-order harmonics must decrease, and eventually it  tends to zero at high frequencies, even with a deep modulation, as is shown in Ref.~\cite{FANTretyakov2020}. For this reason, the bulk-component model and the notion of capacitance, inductance, and resistance can be used also for time-modulated elements of electrically small sizes at all relevant frequencies.

Let us first discuss the modulation by mechanical means, say, by changing the distance between the plates in the absence of material filling. 
The relation between the surface charge density and the normal component of electric flux density (that in this case equals to $\epsilon_0$ multiplied by the normal component of the electric field at the plates) is instantaneous, as it follows from integration of $\nabla\cdot \mathbf{D}=\rho$ over a box of a negligible thickness. But since we have assumed that the electrical size of the capacitor is negligible, we can neglect the time delay between the change of the electric field right at the plates and the corresponding change of the voltage between the plates. Thus, the relation between the charge and voltage can be assumed to be instantaneous, and in this case it is possible to use the conventional definition of capacitance, writing $Q(t)=C(t)V(t)$, where capacitance $C(t)$ depends on time due to external modulation. 

Let us next consider the case when the space between the plates is filled with an electrically polarizable medium whose properties are changing in time due to some external force. In this case, we need to account for time (frequency) dispersion of the filling material. Assuming that the response is linear (and, of course, causal), the polarization density $\mathbf{P}(t)$ between the plates is coupled with the electric field as
\begin{equation}
\mathbf{P}(t)=\epsilon_0\int_0^\infty\chi(\gamma,t)\mathbf{E}(t-\gamma)d\gamma 
\end{equation}
with the real-valued function $\chi(\gamma,t)$ being the susceptibility kernel. Writing the definition of vector $\mathbf{D}$,  $\mathbf{D}(t)=\epsilon_0\mathbf{E}(t)+\mathbf{P}(t)$, we readily conclude that the electric flux density $\mathbf{D}(t)$ is
\begin{equation}
\mathbf{D}(t)=\epsilon_0\int_0^\infty\epsilon_{\rm{k}}(\gamma,t)\mathbf{E}(t-\gamma)d\gamma,
\end{equation}
where 
\begin{equation}
\epsilon_{\rm{k}}(\gamma,t)=\delta(\gamma)+\chi(\gamma,t)
\end{equation}
is the relative permittivity kernel in time domain. The normal component of the electric flux density equals the surface charge density on the plates. By integrating the surface charge density over the surface of the plate, we find the total free charge on the plates. Therefore, by assuming homogeneous charge distribution on the plates, we have the instantaneous charge $Q(t)$ on the plate as
\begin{equation}
Q(t)=A\cdot\epsilon_0\int_0^\infty\epsilon_{\rm{k}}(\gamma,t)\Big[\mathbf{a}_{\rm{n}}\cdot\mathbf{E}(t-\gamma)\Big]d\gamma,
\end{equation}
where $A$ and $\mathbf{a}_{\rm{n}}$ are the area and the unit vector normal to the surface, respectively. 

On the other hand, the integration of $\mathbf{E}(t)$ along a route between the two plates gives the voltage difference $V(t)$ between the two plates. Recalling the assumption of negligible electric size of the capacitor at all relevant frequencies, we use the  quasi-static and homogeneous approximation, where the voltage is simply the multiplication of the normal electric field and the distance $d$ between the two plates. Thus, the above equation can be rewritten as 
\begin{equation}
Q(t)=\int_0^\infty C_{\rm{k}}(\gamma,t)V(t-\gamma)d\gamma,
\label{eq:qtperkvol}
\end{equation}
where
\begin{equation}
    C_{\rm{k}}(\gamma,t)=\epsilon_0\epsilon_{\rm{k}}(\gamma,t){A\over d}\label{eq:Cepsrelation}
\end{equation}
is the capacitance kernel in time domain. This is a general relation that connects the charge on the plates and the voltage difference between them. Here, $C_{\rm{k}}(\gamma,t)$ depends on the observation time $t$, at which we measure the electric charge, and the delay time $\gamma$ between the charge measurement and the applied time-varying voltage difference. 

Now, suppose that the source voltage is time-harmonic, in other words, 
\begin{equation}
V_s(t)=\Re\Big[V_0 e^{j\omega t}\Big],
\end{equation}
where $V_0$ is the complex amplitude, $\omega$ is the angular frequency of the signal, and $\Re[~]$ denotes the real part of the expression inside the bracket. Based on the above two equations, we find that the charge is simply  
\begin{equation}
Q(t)=\Re\Big[C(\omega,t)\cdot V_0 e^{j\omega t}\Big],
\label{eq:qctemvol}
\end{equation}
where
\begin{equation}
C(\omega,t)=\int_0^\infty C_{\rm{k}}(\gamma,t) e^{-j\omega\gamma}d\gamma.\label{C9}
\end{equation}
The right-hand side of \eqref{C9} is the Fourier transform of the capacitance kernel $C_{\rm{k}}(\gamma,t)$ with respect to the time delay variable $\gamma$. Therefore, the function $C(\omega,t)$ on the left-hand side explicitly depends on the angular frequency due to the inherent dispersion of the material filling, in addition to the dependency on time due to the nonstationarity. If there is no time modulation of the material filling the space between the two plates, $C(\omega,t)$ is only a function of the angular frequency and $\partial C(\omega,t)/\partial t=0$. 

We call $C(\omega,t)$   \emph{temporal capacitance}. Similarly, \emph{temporal permittivity} of the filling medium is defined as  
\begin{equation}
\epsilon(\omega,t)=\int_0^\infty\epsilon_{\rm{k}}(\gamma,t) e^{-j\omega\gamma}d\gamma. \end{equation}
In view of Eq.~\eqref{eq:Cepsrelation}, we can easily conclude that for parallel-plate capacitors the temporal capacitance and the temporal permittivity are related in the usual way as 
\begin{equation}
C(\omega,t)=\epsilon_0\epsilon(\omega,t){A\over d}.
\end{equation}

The electric current through the capacitor is the time derivative of the charge, i.e. $I(t)=dQ(t)/dt$. Hence, by employing Eq.~\eqref{eq:qctemvol}, we deduce that  
\begin{equation}
I(t)=\Re\Bigg[\Big(j\omega C(\omega,t)+{\partial C(\omega,t)\over\partial t}\Big)\cdot V_0 e^{j\omega t}\Bigg].\label{cap_def}
\end{equation}
In the stationary case when the capacitor properties are time-invariant, the time derivative of the temporal capacitance is zero, and we get the usual expression: $I_0=j\omega CV_0$ in which $I_0$ is the complex amplitude of the electric current. Let us introduce the following definition: 
\begin{equation}
\begin{split}
C_{\rm{eq}}(\omega,t)&=C(\omega,t)+{1\over j\omega}{\partial C(\omega,t)\over\partial t}\cr
&=\epsilon_0\Big[\epsilon(\omega,t)+{1\over j\omega}{\partial\epsilon(\omega,t)\over\partial t}\Big]{A\over d}.
\end{split}
\end{equation}
By this way, we can use  the conventional expression for the electric current also for time-modulated capacitors, writing 
\begin{equation}
I(t)=\Re\Big[j\omega C_{\rm{eq}}(\omega,t)\cdot V_0 e^{j\omega t}\Big].
\label{cap_def1}
\end{equation}
In other words, we can define \emph{temporal admittance} as
\begin{equation}
Y(\omega,t)=j\omega C_{\rm{eq}}(\omega,t).
\label{eq:admiY}
\end{equation}

Interestingly, the equivalent temporal capacitance $C_{\rm{eq}}(\omega,t)$ is complex-valued rather than a real value. If we assume that the temporal capacitance $C(\omega,t)$ is real within a certain frequency range, the real and imaginary parts of $C_{\rm{eq}}(\omega,t)$ equal to
\begin{equation}
\begin{split}
&\Re\big[C_{\rm{eq}}(\omega,t)\big]=C(\omega,t),\cr
&\Im\big[C_{\rm{eq}}(\omega,t)\big]=-{1\over\omega}{\partial C(\omega,t)\over\partial t}.
\end{split}
\end{equation}
Therefore, the imaginary part of the equivalent temporal capacitance is proportional to the time derivative of the temporal capacitance. Recalling  Eq.~\eqref{eq:admiY}, we find  that the real part of the temporal admittance is $\Re[Y(\omega,t)]=\partial C(\omega,t)/\partial t$. What does this mean? It shows that the time derivative of the temporal capacitance $C(\omega,t)$ can be interpreted as an effective  positive or negative \emph{temporal resistance}. If the time derivative is positive, we have a positive resistance, and there is an effective time-varying negative resistance in the case of negative time derivative.   

Based on the above results, we can use Eqs.~\eqref{cap_def} and \eqref{cap_def1} to find current related to all frequency components of the  voltage. Importantly, because the system is linear, we can use this technique to find response to any periodically-varying voltage. To do that, we expand the voltage function into the Fourier series
 and use the temporal impedance as a function of the frequency. 
 
\subsection{Time-varying inductance}

Similarly to the case of a capacitor, for a time-modulated inductor, the first assumption is that the size of the inductor is electrically small compared to the wavelength at all relevant frequencies. That is, the inductor can be viewed as a bulk component. 
Also in this case, time modulation can be realized either by changing the coil sizes or by changing the magnetic properties of the coil core. In the first case, the relation between the magnetic flux and the current can be assumed to be instantaneous, and we can write $\varphi(t)=L(t)I(t)$. If the core is filled with a dispersive magnetic material, the relation takes the general linear causal form: 
\begin{equation}
\varphi(t)=\int_0^\infty L_{\rm{k}}(\gamma,t)I(t-\gamma)d\gamma,
\end{equation}
where $L_{\rm{k}}(\gamma,t)$ is the inductance kernel in time domain. Using the Fourier transform, we define \emph{temporal inductance} as
\begin{equation}
L(\omega,t)=\int_0^\infty L_{\rm{k}}(\gamma,t) e^{-j\omega\gamma}d\gamma .\label{temp_ind}
\end{equation}
Assuming a time-harmonic current source $I_s(t)=\Re[I_0 e^{j\omega t}]$, we readily conclude that
\begin{equation}
\varphi(t)=\Re\Big[L(\omega,t)\cdot I_0 e^{j\omega t}\Big].\label{flux}
\end{equation}
Since the voltage difference over the element $V(t)$ is the time derivative of the magnetic flux,  after simple algebraic manipulations, we arrive to 
\begin{equation}
V(t)=\Re\Big[j\omega L_{\rm{eq}}(\omega,t)\cdot I_0 e^{j\omega t}\Big], \label{flux_20}
\end{equation}
where the equivalent temporal inductance $L_{\rm{eq}}(\omega,t)$ reads
\begin{equation}
L_{\rm{eq}}(\omega,t)=L(\omega,t)+{1\over j\omega}{\partial L(\omega,t)\over\partial t}.\label{L_eq}
\end{equation}
As it is seen, $L_{\rm{eq}}(\omega,t)$ is complex-valued, and, based on the above two equations, we define the \emph{temporal impedance}  $Z(\omega,t)$ as
\begin{equation}
\begin{split}
Z(\omega,t)&=j\omega L_{\rm{eq}}(\omega,t)\cr
&={\partial L(\omega,t)\over\partial t}+j\omega L(\omega,t).
\end{split}
\end{equation} 
It is clear that positive (negative) temporal resistance is achievable when the time derivative of the temporal inductance is positive (negative).


\section{General tool for analyzing time-modulated elements}
\label{sec:gttmeppp}

As we have shown above, the response of time-modulated circuit components is described by temporal capacitance, inductance, and resistance which depend both on time and the frequency. We have seen that the frequency dependence appears due to frequency dispersion of materials from which the components are made (usually, dielectric fillings of capacitors and magnetic fillings of inductors).  
If the permittivity and permeability of material filling has negligible frequency dispersion in the relevant frequency range, we can consider temporal capacitance, temporal admittance, and temporal resistance as functions of time only. 


Next, we develop a general tool for analyzing electrical systems consisting of periodically time-modulated dispersive elements. The modulation function can be arbitrary, limited only by the assumption of electrically small size of the components. A similar matrix formulation without considering the dispersion effect has been presented in early studies~\cite{Tucker}. Here, we systematically develop the general theory as the foundation of the proposed new applications in this manuscript.
We will first analyze the effect of time-modulated lumped elements, i.e., inductor, capacitor, and resistor, with the voltage-current relation in time domain, and then investigate their effects in circuits. 

\subsection{Time-modulated inductors} \label{sec: inductor}

Let us consider a time-varying inductor, modeled by its temporal inductance \eqref{temp_ind}. If the modulation is periodical with the period $T$, the inductance function satisfies $L(\omega, t)=L(\omega, t+nT)$ with $n \in \mathbb{Z}$.  Therefore, the temporally varying inductance can be expanded into complex Fourier series 
\begin{equation}
    L(\omega, t)=\sum_{p=-\infty}^{+\infty}l_p(\omega) e^{jp\wM t},\label{Eq:inductance_modulation_function}
\end{equation}
\noindent where $\wM=2\pi/T$ is the fundamental angular frequency of modulation, and the complex coefficients $l_p(\omega)=\frac{1}{T}\int_0^T L(\omega, t)e^{-jp\wM t}dt$ are the modulation coefficients (amplitudes and phases) at angular frequencies $p\wM$. 
If losses in the inductor material can be neglected, $L(\omega, t)$ is a real-valued function, and the modulation coefficients satisfy the condition $l_p(\omega)=l^*_{-p}(\omega)$. In the general case, $L(\omega,t)$ is a complex-valued causal response function, which satisfies the Kramers-Kronig relation at every value of the time argument $t$ \cite{Engheta2020KKR}. 

Let us assume that a periodically time-modulated inductor is connected to an electrical circuit which is excited by external sources. According to  the Floquet theorem, if the external excitation has a  time-harmonic component $e^{j\ws t}$ at the angular frequency $\ws$, the corresponding  voltage $V(t)$ and current $I(t)$ across this time-modulated inductor will exhibit an  infinite number of Floquet harmonics, and they can be expressed by  \cite{XCWangthesis,Elnaggarmodeling}
\begin{equation}
    	V(t)  =  \sum_{n=-\infty}^{+\infty}v_n e^{j\wn t},\label{Eq:VL(t)}
\end{equation}
and 
\begin{equation}
    	I(t)  =  \sum_{n=-\infty}^{+\infty}i_n e^{j\wn t}\label{Eq:iL(t)},
\end{equation}
where $v_n$ and $i_n$ are  complex-valued coefficients, and
\begin{equation}
	\wn=\ws+n\wM. \label{Eq:wn}
\end{equation}
Note that in Eqs.~(\ref{Eq:VL(t)}) and (\ref{Eq:iL(t)}), we omit the operator $\Re$ in front of the summation sign, for the mathematical convenience in the following derivations.

If the excitation contains several time-harmonic components at different angular frequencies, 
the voltage across (and current through) the time-modulated inductor will be a superposition of Floquet harmonics of Eqs.~(\ref{Eq:VL(t)}) and (\ref{Eq:iL(t)}) for multiple frequency sets in Eq.~(\ref{Eq:wn}), as follows from the linearity of the circuit. 


For each frequency harmonic of the current, the corresponding voltage is given by Eq.~\eqref{flux_20}. Let us consider the $n$-th current harmonic, $I_n(t)=i_n e^{j\omega_nt}$. For this harmonic, the corresponding voltage across the time-varying inductor can be obtained by substituting $I_n(t)$ and the Fourier expansion of the inductance \eqref{Eq:inductance_modulation_function} into (\ref{flux_20}), which gives 
\begin{equation}
    V_n(t)=\sum_{p}j(\omega_n+p\omega_{\rm M})l_p(\omega_n)e^{jp\omega_{\rm M}t}\cdot i_n e^{j\wn t}.
\end{equation}
Because of the linearity, we can find the total voltage [see Eq.~\eqref{Eq:VL(t)}] by summing up the  contributions induced by all the  frequency harmonics of the current,
\begin{equation}
 \sum_{n}v_n e^{j\omega_nt}=\sum_{n}\sum_{p}j\omega_{n+p}l_p(\omega_n)i_ne^{j\omega_{n+p}t},\label{eq: LI substituted}
\end{equation}
\noindent where the summation over $n$ and $p$ is from $-\infty$ to $+\infty$.
By replacing the index $n$ by $n-p$ at the right side of Eq.~(\ref{eq: LI substituted}), we have
\begin{equation}
 \sum_{n}v_n e^{j\omega_nt}=\sum_{n}\sum_{p}j\omega_nl_p(\omega_{n-p})i_{n-p}e^{j\omega_nt}.\label{eq: LI substituted1}
\end{equation}
Now we can see that Eq.~(\ref{eq: LI substituted1}) shares the same basis on both sides, and we can equate the corresponding coefficients:
\begin{equation}
    v_n=\sum_{p}j\omega_nl_p(\omega_{n-p})i_{n-p}, \label{Eq:harmonic relation V_L}
\end{equation}
where $p$ is from $-\infty$ to $+\infty$.
From this equation we clearly see that when the inductor is modulated in time, the voltage component at the frequency $\wn$ is not only related to the current at the same frequency $\wn$, but also to  the currents at other frequencies $\omega_{n-p}$, with the coefficients $j\wn l_p(\omega_{n-p})$ playing the role of mutual impedances. 

Since all the voltage harmonics satisfy relation (\ref{Eq:harmonic relation V_L}), we can write this  linear equation in a matrix form:
\begin{equation}
	{\bf v}=\ZL\cdot{\bf i}, \label{Eq:VI_f_inductor}
\end{equation}
where vectors ${\bf v}={\{v_{-\infty},...,v_{-1},v_0,v_1,...,v_{\infty}\}}$ and ${\bf i}={\{i_{-\infty},...,i_{-1},i_0,i_1,...,i_{\infty}\}}$ resemble the complex spectra of the voltage and current at frequencies ${\{\omega_{-\infty},...,\omega_{-1},\omega_{\rm s},\omega_1,...,\omega_{\infty}\}}$.
The matrix $\ZL$ is therefore the impedance matrix that relates the voltage and current in the frequency domain. This impedance matrix contains off-diagonal terms due to the mode coupling introduced by  time modulation.
	
Usually, the modulation depth is relatively small and the mode coupling mainly takes place at frequencies close to $\ws$.  In this case, we can simplify the system and consider only the harmonics with $n$ from $-N$ to $+N$ with the truncation index $N$ being a finite integer. 
In fact, it is necessary to limit the order of harmonics in order to ensure that we stay within the assumption of electrically small, bulk components. Since the component size has been assumed to be much smaller than the wavelength at all relevant frequencies, the truncation index must satisfy
\begin{equation}
    N \ll {c\over D f_s},
    \label{c_bound}
\end{equation}
where $c$ is the speed of light, $D$ is the size of the component, and $f_s=\omega_{\rm s}/2\pi$ is the fundamental frequency. In practice, the lower bound of the truncation number $N$ that is defined by  the strengths of cross-coupling between higher-order harmonics, is considerably smaller than the bound given by \eqref{c_bound}.
This practical bound depends mainly on the  modulation depth, as we will discuss  in Section~\ref{sec:tmiprlc}.

Thus,  the voltage and current vectors become finite-dimension $2N+1$ vectors. The impedance matrix in Eq.~(\ref{Eq:VI_f_inductor}) is a $2N+1$ by $2N+1$ matrix instead of an infinitely large one, and it reads
\begin{equation}
\newcommand\scalemath[2]{\scalebox{#1}{\mbox{\ensuremath{\displaystyle #2}}}}
	\ZL=j\Wmat\scalemath{0.905}{\begin{pmatrix}
	l_0(\omega_{-N}) 	& l_{-1} (\omega_{1-N})	& \cdots & l_{-2N} (\omega_{N}) \\
	l_1(\omega_{-N})	& l_{0} (\omega_{1-N})	& \cdots & l_{1-2N}(\omega_{N}) \\
	\vdots  & \vdots 	& \ddots & \vdots  \\
	l_{2N}(\omega_{-N})	& l_{2N-1}(\omega_{1-N})	& \cdots & l_0(\omega_{N})
	\end{pmatrix}}, \label{Eq:ZLmat_full}
\end{equation}
where
\begin{equation}
	\Wmat={\rm diag}\{\omega_{-N},...,\omega_{-1},\omega_{\rm s},\omega_1,...\omega_N\},\label{Eq:Wmat}
\end{equation}
is a diagonal matrix with the  elements being the mode frequencies defined in Eq.~(\ref{Eq:wn}). We note that this impedance matrix reduces to the conventional scalar model when the inductor is not modulated in time and is  dispersionless. Without modulation and frequency dispersion, the inductance is a constant value $L_0$ and, therefore, the coefficients in Eq.~(\ref{Eq:inductance_modulation_function}) simplify to the Kronecker delta function:  $l_p=\delta_{p}L_0$. As a result, the above impedance matrix is a diagonal one, and the voltage at frequency $\ws$ is related only with the current at the same frequency by impedance $j\ws L_0$.

\subsection{Time-modulated capacitors}
Similarly, for a periodically time-modulated dispersive capacitor, we can do the same to characterize it in the frequency domain. However, instead of the  impedance matrix, it is much more convenient to use the admittance matrix, considering the current as a function of voltage (the current is the time derivative of the multiplication of capacitance and voltage). If we assume that the capacitance is modulated with the same period $T$, then it can be expressed in Fourier series 
\begin{equation}
	C(\omega, t)=\sum_{p=-\infty}^{+\infty}c_p(\omega) e^{jp\omega_{\rm M} t}, \label{Eq:capacitor_modulation_function}
\end{equation}
where $c_p(\omega)=\frac{1}{T}\int_0^T C(\omega,t)e^{-jp\wM t}dt$ are the complex modulation coefficients which depend on the frequency. Then, under an external excitation with a time-harmonic component $e^{j\ws t}$ at the angular frequency $\ws$, the voltages/currents across/through the capacitor can be expressed in the same way as in Eqs.~(\ref{Eq:VL(t)}) and~(\ref{Eq:iL(t)}). Using the relation in \eqref{cap_def1} and following the same derivation procedure as in Sec.~\ref{sec: inductor},
we get a similar matrix voltage-current relation in the frequency domain:
\begin{equation}
	{\bf i}=\YC\cdot{\bf v}, \label{Eq:VI_f_capacitor}
\end{equation}
where $\YC$ is the admittance matrix for the time-modulated capacitor. Considering a finite number of harmonics with $-N\leq n\leq+N$, it can be written as
\begin{equation}
\newcommand\scalemath[2]{\scalebox{#1}{\mbox{\ensuremath{\displaystyle #2}}}}
	\YC=j\Wmat\scalemath{0.885}{\begin{pmatrix}
	c_0(\omega_{-N}) 	& c_{-1} (\omega_{1-N})	& \cdots & c_{-2N} (\omega_{N}) \\
	c_1(\omega_{-N})	& c_{0} (\omega_{1-N})	& \cdots & c_{1-2N}(\omega_{N}) \\
	\vdots  & \vdots 	& \ddots & \vdots  \\
	c_{2N}(\omega_{-N})	& c_{2N-1}(\omega_{1-N})	& \cdots & c_0(\omega_{N})
	\end{pmatrix}},\label{Eq:YCmat_full}
\end{equation}
which is similar to Eq.~(\ref{Eq:ZLmat_full}). We also note that when the capacitor has a constant capacitance $C_0$ at all frequencies (no time modulation and frequency dispersion), the admittance matrix reduces to a  diagonal one with the elements $j\ws C_0$.

\subsection{Time-modulated resistor}
Finally, for a periodically time-modulated resistor, the treatment is easier as the time-domain V-I relation is much simpler, i.e., $V(t)=R(\omega, t)I(t)$.
Following the same procedure and assuming that the periodically time-modulated resistance is expanded as
\begin{equation}
	R(\omega,t)=\sum_{p=-\infty}^{+\infty}r_p(\omega) e^{jp\wM t}, \label{Eq:resistance_modulation_function}
\end{equation}
the impedance matrix of a time-modulated resistor is obtained as
\begin{equation}
\newcommand\scalemath[2]{\scalebox{#1}{\mbox{\ensuremath{\displaystyle #2}}}}
	\ZR=\scalemath{0.955}{\begin{pmatrix}
	r_0(\omega_{-N}) 	& r_{-1} (\omega_{1-N})	& \cdots & r_{-2N} (\omega_{N}) \\
	r_1(\omega_{-N})	& r_{0} (\omega_{1-N})	& \cdots & r_{1-2N}(\omega_{N}) \\
	\vdots  & \vdots 	& \ddots & \vdots  \\
	r_{2N}(\omega_{-N})	& r_{2N-1}(\omega_{1-N})	& \cdots & r_0(\omega_{N}) 
	\end{pmatrix}},\label{Eq: Zr}
\end{equation}
satisfying ${\bf v}=\ZR\cdot{\bf i}$. We note that the modulation coefficients $r_p(\omega)$ are complex numbers, meaning that the time-modulated resistance can produce an effective reactance. When the resistance is time-invariant and non-dispersive, the impedance matrix reduces to a diagonal one with the same diagonal value $r_0=R_0$.

Now, we can characterize time-modulated elements with impedance or admittance matrices in frequency domain. This is a powerful tool to analyze any electrical circuit that contains periodically-modulated elements. In the following, we consider a few important examples to verify the developed tool on  practically relevant examples of novel usage of time-modulated components. 


\section{Time-modulated inductor in an $RLC$-circuit}
\label{sec:tmiprlc}

\begin{figure}[!b]
	\centerline{\includegraphics[width= \columnwidth]{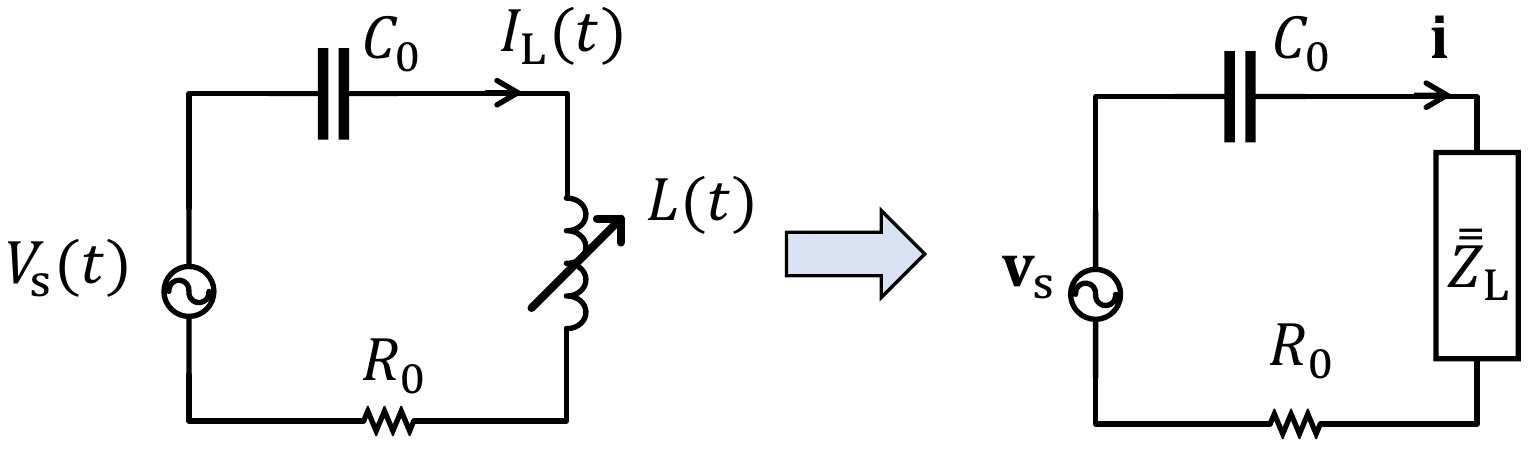}}
	\caption{An $RLC$-circuit with a time-modulated inductor and its equivalent circuit representation.}
	\vspace{-10pt}
	\label{Fig:RL(t)C_circuit}
\end{figure}

Let us first consider an $RLC$-circuit driven by a voltage source at frequency $\ws$, 
\begin{equation}
	V_s(t)=\Re{[V_0 e^{j\ws t}]}, \label{Eq:Vs(t)}
\end{equation}
with $L$ being time-modulated. The circuit under study is shown in Fig.~\ref{Fig:RL(t)C_circuit} at the left. Let us assume that the inductor is dispersive and modulated in a harmonic way around its nominal values $L_0(\omega)$ at frequency $\wM$ with the same modulation depth $m_{\rm M}$: 
\begin{equation} \label{Eq:L(t)cos}
	L(\omega, t)=L_0(\omega)\Big(1 + \mM \cos(\wM t+\phiM)\Big),
\end{equation}
where the variables $\mM$, $\phiM$, and $\wM$ indicate the modulation depth, phase, and frequency, respectively (this convention is used throughout the paper). Similarly to Eq.~(\ref{Eq:inductance_modulation_function}), we write this function in the exponential form
\begin{equation} \label{Eq:L(t)exp}
	L(\omega, t) =l_0(\omega)+ l_{-1}(\omega) e^{-j\wM t}+ l_1(\omega) e^{j\wM t},
\end{equation}
where $l_0=L_0(\omega)$, and $l_{\pm1}(\omega)=\frac{1}{2}\mM L_0(\omega) e^{\pm j\phiM}$ are the complex modulation coefficients at the modulation frequencies $\pm\wM$, respectively. 

\subsection{The impedance matrices and the master equation}
For this simple time-modulated inductor, we find from Eq.~(\ref{Eq:ZLmat_full}) that its impedance matrix  is a $2N+1$ by $2N+1$ tri-diagonal matrix,
\begin{equation} 
\newcommand\scalemath[2]{\scalebox{#1}{\mbox{\ensuremath{\displaystyle #2}}}}
	\ZL=j\Wmat\scalemath{0.835}{\begin{pmatrix}
		l_0(\omega_{-N})		& l_{-1}(\omega_{1-N}) & 0 		& \cdots & 0\\
		l_1(\omega_{-N})		& l_0 (\omega_{1-N})	 & l_{-1}(\omega_{2-N})	& \cdots & 0 \\
		0     	& l_1(\omega_{1-N})  	 & l_0(\omega_{2-N}) 		& \ddots & 0 \\
		\vdots	& \vdots & \ddots   & \ddots & \vdots  \\
		0      	& 0      & 0  		& \cdots & l_0(\omega_{N})
	\end{pmatrix}}. \label{Eq:ZLmat_param_amp}
\end{equation}
We can see that frequency component coupling takes place only at the same and nearest frequencies. 

In general, all the frequencies  are different, i.e., $\wn$ and $\omega_{n+1}$ have different (absolute) values for all $n$. 
However, when the modulation frequency is $\wM=2\ws$, there are two neighbor components with the same frequency, i.e., $\ws$ and $\omega_{-1}=-\ws$, meaning that there will be very strong effect at the signal frequency $\ws$ due to the time modulation of the inductor, which is the same effect as used in classical parametric amplifiers~\cite{Suhl}. Let us  derive the master equation of this circuit. 

As the $RC$ components are dispersive but not time-modulated, matrices $\ZR$ and $\YC$ can be obtained by replacing their off-diagonal terms in Eqs.~(\ref{Eq: Zr}) and (\ref{Eq:YCmat_full}) with zeros.
 Therefore, the master equation of the $RLC$-circuit in frequency domain is
\begin{equation}
	{\bf v}_s=(\ZR+\ZL+\YC^{-1})\cdot{\bf i}, \label{Eq:RLC-master}
\end{equation}
where ${\bf v}_s$ and $\bf i$ are the source voltage vector and the current vector, respectively. For  the source voltage in Eq.~(\ref{Eq:Vs(t)}), the voltage vector ${\bf v}_s$ has components $V_0/2$ at frequencies $\pm\ws$ and 0 at all other frequencies. Therefore, we can easily calculate the current vector from the master equation (\ref{Eq:RLC-master}) and then obtain the current in time domain by taking the real part of Eq.~(\ref{Eq:iL(t)}). The voltages across the elements can be also obtained similarly. 

Here, we conclude by formulating a simple recipe for establishing the master equation for general linear electrical circuits consisting of dispersionless time-modulated elements: 1. write down the $V-I$ equation in frequency domain assuming that  there is no time-modulated element; 2. expand the frequency spectrum based on Eqs.~\eqref{Eq:wn} and \eqref{Eq:Wmat}; 3. for each electrical component including the time-modulated one,  replace the scalar impedance by the corresponding  matrix impedance, as is illustrated in Fig.~\ref{Fig:RL(t)C_circuit}. Next, we discuss this particular procedure in detail in both time and frequency domains. 

\subsection{Time-domain analysis}
Let us consider a particular numerical example to verify the above theoretical discussion (we refer to the same $RLC$-circuit shown in Fig.~\ref{Fig:RL(t)C_circuit}). In subsequent numerical examples we assume that the time-modulated components are dispersionless. In this example, we assume that the voltage source is defined by  Eq.~(\ref{Eq:Vs(t)}) at $100$~kHz, i.e., $\ws=2\pi\times100~{\rm krad/s}$, with the  magnitude $V_0=1~{\rm V/m}$, and the inductor is modulated as shown in Eq.~(\ref{Eq:L(t)cos}) at the double frequency $\wM=2\ws$ around the inductance value $L_0=100~{\rm \mu H}$ (note that in this example, $L_0$ is not frequency dependent because we neglect frequency dispersion of its materials). In addition, we choose $R_0=100~{\rm \Omega}$ and $C_0=1/(\ws^2 L_0)$, so that the circuit is resonant at the signal frequency $\ws$ when the inductor is not modulated. Then, the modulation depth $\mM$ and modulation phase $\phiM$ are varied to investigate the circuit performance.

 \begin{figure}[!b]
	\centerline{\includegraphics[width= \columnwidth]{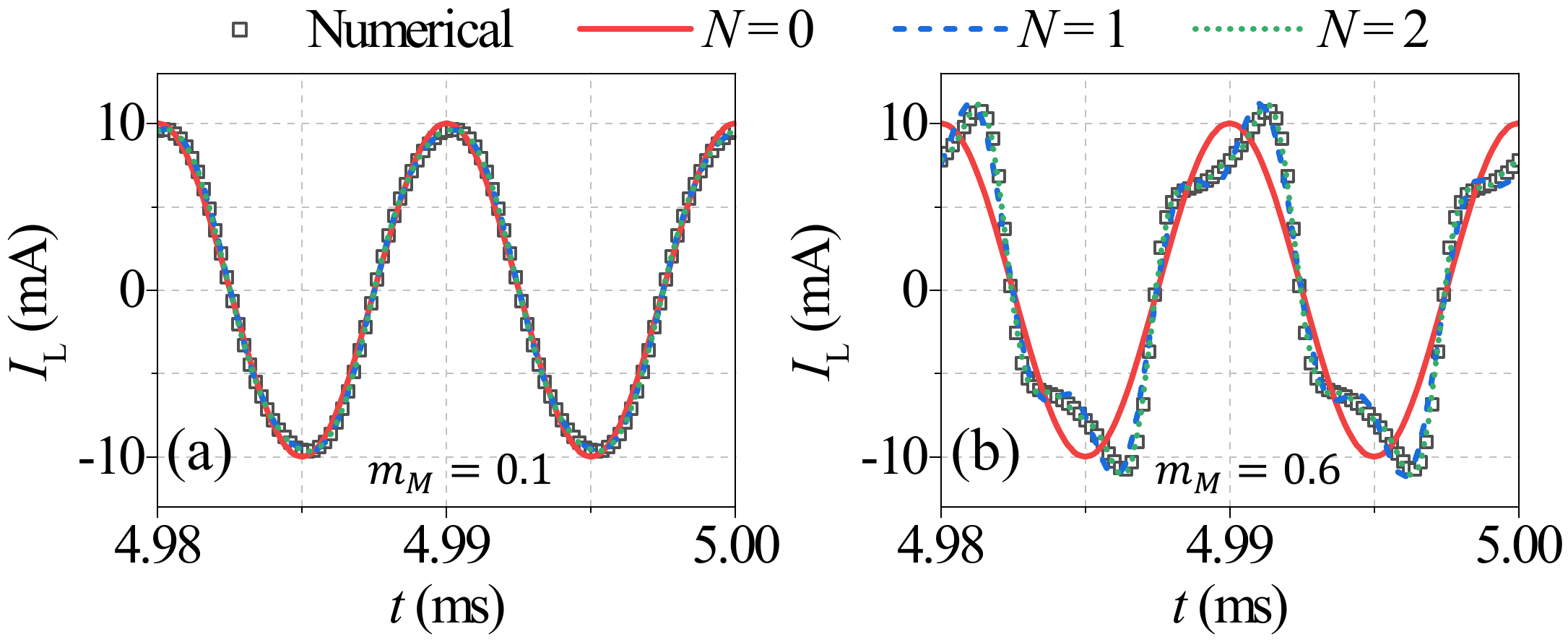}}
 		\caption{Numerical simulation of the current waveform of $I(t)$ for (a)~modulation depth $\mM=0.1$ and (b)~modulation depth $\mM=0.6$. Simulation is performed for $\phiM=0$, $R_0=100~{\rm \Omega}$, $L_0=100~{\rm \mu H}$, and $\omega_s=2\pi\times100~{\rm krad/s}$.}
 		\vspace{-10pt}
 		\label{Fig:I_L_Waveform}
   \end{figure}

The developed analytical tool is validated by comparing  time-domain solutions. First,  the  $RLC$ circuit can be numerically studied by solving the corresponding  time-domain differential equations. This is done using \emph{Mathematica} software,  and the black open square markers in Fig.~\ref{Fig:I_L_Waveform} show the numerically calculated time-domain steady state current waveforms $I(t)$. Then, we analyze the performance with the proposed tool and compare the  results. In this case, the choice of the truncation index $N$ matters. For example, when the modulation depth is small (with $\mM=0.1$ and $\phiM=0~{\rm rad}$), a small truncation index is enough to represent the circuit performance, as shown by the red line with $N=0$ in Fig.~\ref{Fig:I_L_Waveform}(a). This is because the coupling between the harmonics is negligible due to the relatively small $l_{\pm}$. However, when the modulation becomes stronger (e.g. $\mM=0.6$ with $\phiM=0~{\rm rad}$), the coupling between harmonics become stronger as $l_{\pm}$ are of the same order as $l_0$, and we should use a  larger truncation index. This is illustrated in  Fig.~\ref{Fig:I_L_Waveform}(b), where we see that the use of a larger index $N=2$ gives an  almost identical current waveform to the numerical one. 
From the time-domain current waveform, it is also clear that the double-frequency modulation (i.e. $\wM=2\ws$) can result in higher-order-harmonic voltages/currents across/through the circuit due to coupling between different modes.


\subsection{Effective impedance of the time-modulated inductor at the signal frequency}
If we know the voltage across and current through a  time-modulated inductor, denoted by $\vLt$ and $\iLt$, respectively, the effective impedance of this time-modulated inductor at the signal frequency $\ws$ can be defined as 
\begin{equation} \label{Eq:ZL_eqv_def)}
    \ZLeff= \dfrac{\mathcal{F} [\vLt]\big|_{\omega=\omega_s}}{\mathcal{F} [\iLt]\big|_{\omega=\ws}} ,
\end{equation}
where $\mathcal{F}[~]$ denotes the Fourier transform.  Note that this definition is for a dispersionless component. Here, the time-modulated element is not directly connected to a monochromatic source but through other electrical elements. Therefore, the above definition is different from the general definition in Section II where we also took into account frequency dispersion. 
In this case, if we assume that the current through the element is simply $\iLt=\cos(\ws t+\theta)$ with the phase angle $\theta$ shift with respect to the signal voltage $\Vst$, then, according to Eq.~(\ref{Eq:VI_f_inductor}), the voltage across the time-modulated inductor contains multiple frequency harmonics. However, the effective impedance $\ZLeff$ at frequency $\ws$ can be found from Eq.~(\ref{Eq:ZL_eqv_def)}) as
\begin{equation} \label{Eq:ZL_eqAna)}
    \ZLeff = j\ws L_0 \Big( 1 + \frac{1}{2} \mM e^{j(\phiM-2\theta)}\Big).
\end{equation}
This is the first-order approximation result, as the multiple voltage harmonics will induce multiple current harmonics and so on so forth. But it already gives useful information on the system. From this result, we can easily see that the time modulation of the inductor contributes additional effective impedance of $\frac{1}{2} j\ws L_0\mM e^{j(\phiM-2\theta)}$. The corresponding first-order approximation for effective resistance $\RLeff$ and effective inductance $\LLeff$ at $\ws$ read
\begin{eqnarray}
	  \RLeff & = & \frac{1}{2} \mM \ws L_0 \sin(2\theta -\phiM),\label{Eq:Reff_L(t)}  \\
	  \LLeff & = & L_0+\frac{1}{2} \mM L_0 \cos(2\theta -\phiM). \label{Eq:Leff_L(t)}    
 \end{eqnarray}

We can see that when the modulation phase $\phiM$ (relative  to angle $\theta$) is varying, the time-modulated inductor exhibits different effective behavior: when $2\theta-\phiM=n\pi~(n \in \mathbb{Z})$, the time modulation effectively contributes additional inductance to $L_0$; while when $2\theta-\phiM=(n+\frac{1}{2})\pi~(n \in \mathbb{Z})$, the time modulation effectively adds positive/negative resistance; for other phase values, the effective resistance varies between $\pm \frac{1}{2} \mM \ws L_0$ and the effective inductance varies between $(1\pm \frac{1}{2} \mM)L_0$, respectively. 

This is an interesting result as compared to the classical literature of parametric amplifiers~\cite{Tien,RayleighParamAmp}  \cite{Collin} and recent advancements in Floquet impedance matching \cite{TM-BandWidth}, where time-modulated elements are only explored as effective negative resistors. This is because usually only a particular phase of the modulation is used and therefore only negative resistance is present.

\begin{figure}[!t]
	\centerline{\includegraphics[width= \columnwidth]{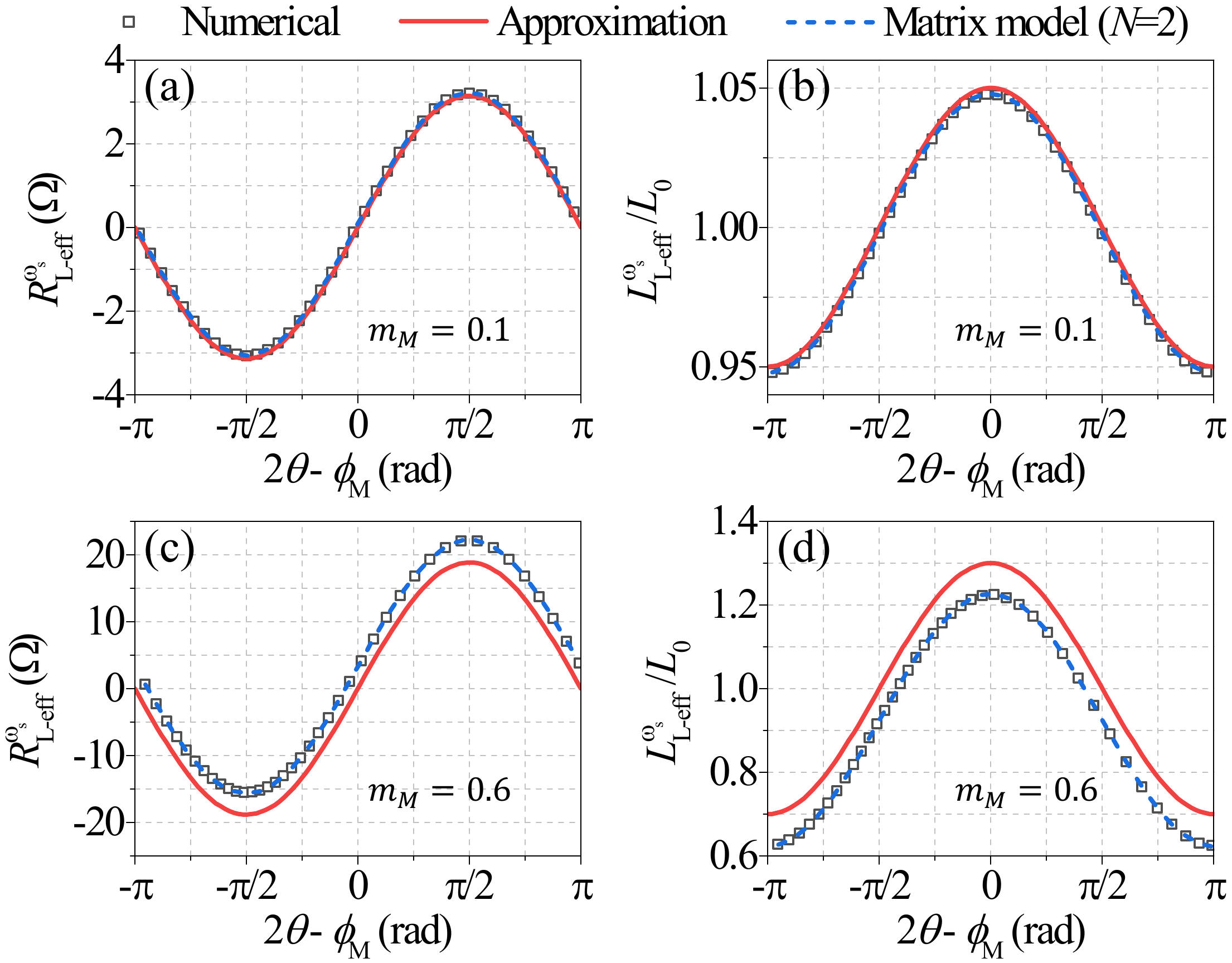}}
 		\caption{The effective resistance and inductance (normalized) of a time-modulated inductor versus $2\theta-\phiM$ for modulation phase $\mM=0.1$ (a, b), and $\mM=0.6$ (c, d). Numerical results are calculated by numerically solving the time-domain differential equations, first-order approximation results are calculated using Eqs. \eqref{Eq:Reff_L(t)} and \eqref{Eq:Leff_L(t)}, and matrix model results are calculated from Eqs. \eqref{Eq:RLC-master} and \eqref{Eq:ZL_eqv_def)}. }
 		\vspace{-10pt}
 		\label{Fig:Re_and_Le_RL(t)CvsPhase}
\end{figure}

Variations of the effective resistance and inductance versus $\phiM$ are evaluated, and the results are given in Fig.~\ref{Fig:Re_and_Le_RL(t)CvsPhase}. We can see that when $\mM=0.1$, the first-order approximation results (red lines) from Eqs. \eqref{Eq:Reff_L(t)} and \eqref{Eq:Leff_L(t)} agree with the  numerical results (black open squares) very well. This is because for low modulation depths the cross coupling between neighbouring harmonics are negligible and the first-order approximation is enough for characterizing the element. However, for large modulation depths, for example $\mM=0.6$, due to strong coupling between close-frequency modes, the induced current at signal frequency from other voltage harmonics are missing in the first-order approximation. Therefore, the results are shifted. Nevertheless, the shape is retained and the range given by the first-order approximation is still valid for large modulation depths. On the other hand, from Fig.~\ref{Fig:Re_and_Le_RL(t)CvsPhase}, we can see that the effective resistance and inductance obtained from the matrix model, i.e., calculated from Eqs. \eqref{Eq:RLC-master} and \eqref{Eq:ZL_eqv_def)} (blue dashed lines), agree exactly with the numerical results. This means that the correct way for modeling the time-varying elements is the impedance matrix method.





\section{Time-modulated mutual inductance for enhanced wireless power transfer}
\label{sec:wpt}
\label{sec:wpttmmi}

In this section, we apply the developed above general analysis tool to time-modulated wireless power transfer (WPT) systems. 
We start with a discussion of the theoretical limits in classical WPT systems. Next, we introduce and discuss a possibility to overcome those fundamental limits using  time-modulated mutual inductance. 

\subsection{Fundamental limitations of wireless power transfer}

Here, we consider classical WPT systems without any time-modulated components.
In these systems, the power transfer capability and the system efficiency are the most important performance indicators. 

\begin{figure}[!t]
	\centerline{\includegraphics[width= 0.7\columnwidth]{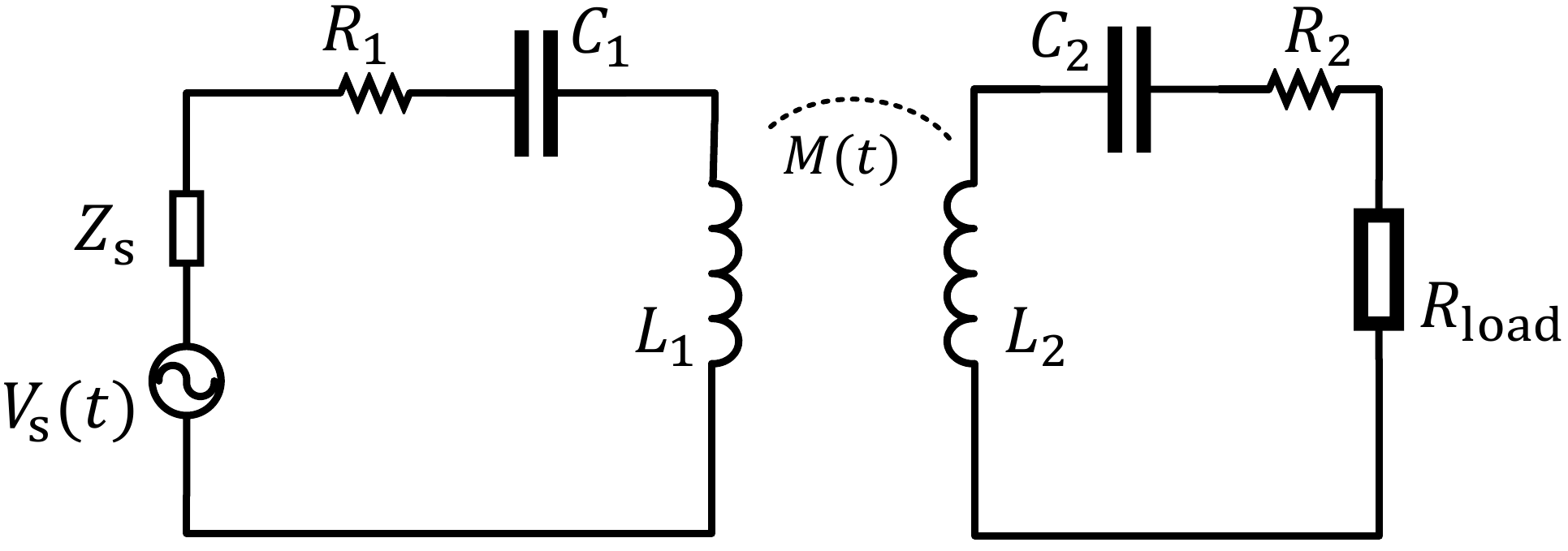}}
 		\caption{Circuit diagram of an inductively coupled WPT system where the mutual inductance can be modulated in  time.}
 		\vspace{-10pt}
 		\label{Fig:WPT_M(t)_circuit}
\end{figure}

A circuit diagram of an inductively coupled WPT system is shown in Fig.~\ref{Fig:WPT_M(t)_circuit}, where we assume that there is no time modulation. The device consists of a harmonic power source $\Vst$ [Eq.~(\ref{Eq:Vs(t)})] with internal impedance $\Zs$, a transmitting resonator (illustrated as an  $LCR$-circuit $L_1$, $C_1$, and $R_1$), and a receiving resonator ($L_2$, $C_2$, and $R_2$) connected to an electrical load $\Rload$. The transmitting resonator inductively couples with the receiving resonator with a mutual inductance of $ {M}$.  
The performance indices, the delivered  power and efficiency will strongly depend on  $ {M}$ and $\Rload$. The input impedance $\Zin$ seen by the source at the resonance frequency $\ws$ is $\Zin=R_1+(\ws  {M})^2/(R_2+ \Rload)$ \cite{WPTref}. The maximum power transfer from the source occurs when $\Zin=\Zs^*$ ($*$ denotes complex conjugation),  which corresponds to a particular $ {M}$  value for a given $\Rload$. Whenever the mutual inductance differs from this value, power transfer is not optimal due to the impedance mismatch in the source terminal. In order to evaluate the transferred power, we define the transferred power ratio $\Rhload$ as the ratio between the output power at the load $\Pload$ and the maximum available power from the source \cite{WPTref}:
\begin{equation}\label{'Eq:TPRdef'}
    \Rhload=\Pload / P_0, 
\end{equation}
where $P_0=V_0^2/(8\Re[\Zs])$ is the maximum available power (used as reference power). On the other hand, power transfer efficiency ($\eta$) is mainly determined by dissipation in the lossy components and the power source, which can be defined as
\begin{equation} \label{Eq:WPTpte}
    \eta  = \dfrac{\Pload }{\PRt+\PRr+\Ps+\Pload},
\end{equation}
\noindent where $\PRt$, $\PRr$, and $\Ps$ are the power losses in the coil resistance $(R_1, R_2)$ and the source resistance $(\Re[\Zs])$.
When the input impedance of the WPT system is matched to the source impedance (i.e., $\Zin=\Zs^*$), the maximum available power from the source is delivered to the WPT system, however, half of the generated power is dissipated inside the source resulting in the efficiency smaller than $50\%$. Therefore, there is always a trade-off between maximizing delivered power and  efficiency.

\subsection{Time-modulated mutual inductance}

Here, we introduce a new possibility of using time-modulated mutual inductance to overcome this fundamental limitation of inductive WPT systems.  We start our discussion by extending the matrix model of time-modulated inductors to modeling of time-modulated mutual inductances. As we have reviewed  in previous sections, a time-modulated inductor or capacitor can equivalently act as a negative resistor. We can introduce time modulation in the transmitting resonator (i.e. $L_1$ or $C_1$) to make the input impedance to be the negative of the source impedance $\Zin=-\Zs$  so that the total impedance in the source loop can be nullified.  In this way, a practical source with internal impedance $\Zs$ is equivalently acting as an ideal source theoretically capable of providing infinite power. Thus, we can use time modulation of $L_1$ or $C_1$ to overcome the theoretical limit of the transferred power which is due to the internal source impedance.  However, because of the physical presence of the source impedance and winding resistances, efficiency of such system will still suffer from the theoretical upper bound discussed above.

On the other hand, time modulation of mutual inductance allows us to overcome physical limitations on both the efficiency and transferred power, as we can directly pump the energy to the receiving resonator without sacrificing the efficiency due to  losses in the primary source. 

\begin{figure}[!b]
	\centerline{\includegraphics[width= \columnwidth]{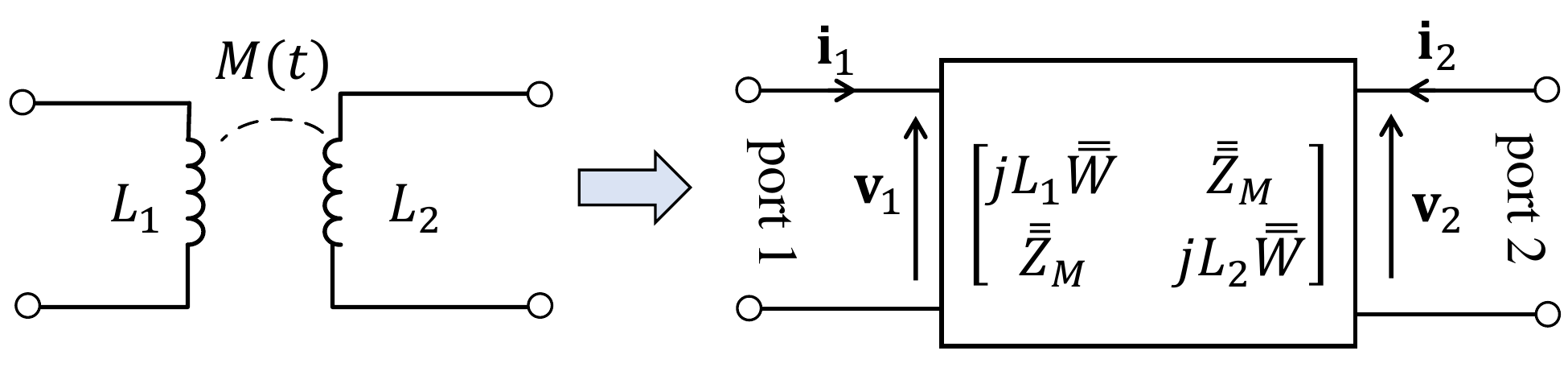}}
 		\caption{Time-modulated mutual inductance and its equivalent circuit.}
 		\vspace{-10pt}
 		\label{Fig:M(t)_circuit}
\end{figure}

Let us consider a time-modulated mutual inductance  in an electrical circuit such as an inductive WPT system. The derivations are similar to the case of time-varying  $L(t)$ in Section~\ref{sec:tmiprlc}, but now we have a two-port network model, as illustrated in Fig.~\ref{Fig:M(t)_circuit}. Let us assume that the mutual inductance $ {M}(t)$  is periodically varied as
\begin{equation} 
 {M}(t)= {M}_0\Big(1 + m_{\rm M} \cos(\wM t+\phi_{\rm{M}})\Big),
\end{equation}
\noindent where $ {M}_0$ is the  nominal mutual inductance,   and  $m_{\rm M}$ is the modulation depth.
Similar to Eq.~\eqref{Eq:L(t)exp}, it is convenient to express $ {M}(t)$ as
\begin{equation} \label{Eq:M(t)}
  {M}(t) ={m}_0+ {m}_{-1} e^{-j  \wM t}+ {m}_1 e^{j  \wM t}
\end{equation}
with ${m}_0= {M}_0$ and ${m}_{\pm1}=\frac{1}{2} m_{\rm M}  {M}_0 e^{\pm j \phi_{\rm{M}}}$.

Using the introduced method of matrix circuit parameters,  we can build a two-port model of the time-modulated mutual inductance as 
\begin{equation}
    \begin{pmatrix} 
        {\bf v}_1\\ {\bf v}_2\\  
    \end{pmatrix} 
    =\begin{pmatrix} 
        jL_1\Wmat & \ZM \\  
        \ZM & jL_2\Wmat  \\  
    \end{pmatrix} 
    \begin{pmatrix} 
        {\bf i}_1\\ {\bf i}_2\\  
    \end{pmatrix},
\end{equation}	
where ${\bf v}_{1,2}$ (${\bf i}_{1,2}$) are the complex spectra of voltage (current) across (through) port~1 (port~2), as illustrated in Fig.~\ref{Fig:M(t)_circuit}. 
Using the same approach as in Section~\ref{sec:gttmeppp}, we can find the impedance matrix $\ZM$ representing the mutual coupling in frequency domain as
\begin{equation} \label{Eq:ZLmatrix}
	\ZM =  j\Wmat\cdot
	\begin{pmatrix}
		{m}_0		& {m}_{-1} & 0 		& \cdots & 0\\
		{m}_1		& {m}_0 	 & {m}_{-1}	& \cdots & 0 \\
		0     	& {m}_1  	 & {m}_0 		& \ddots & 0 \\
		\vdots	& \vdots & \ddots   & \ddots & \vdots  \\
		0      	& 0      & 0  		& \cdots & {m}_0
	\end{pmatrix}.
\end{equation}
Now we can use the matrix model to analyse any electrical circuit with the time-modulated mutual inductance including WPT systems. Our analysis show that the matrix model agrees well with the simulation results when proper $N$ is used. 

\subsection{WPT with time-modulated mutual inductance}

\begin{figure}[!t]
	\centerline{\includegraphics[width= 0.5\columnwidth]{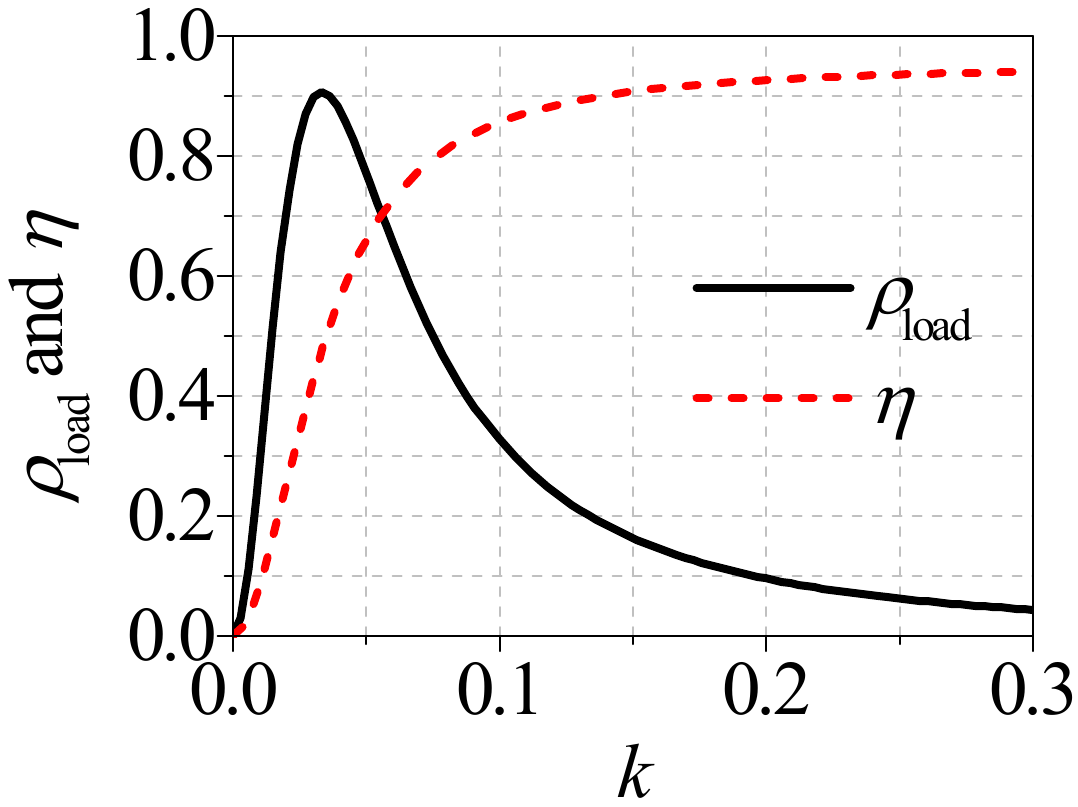}}
 		\caption{The variations of transferred power ratio $\Rhload$ and power transfer efficiency $\eta$ with respect to the coupling coefficient $k$ in a classical unmodulated WPT system.}
 		\label{Fig:WPT_PTE_TPR_vscoupling}
\end{figure}

To demonstrate the capabilities of time-modulated mutual inductance, let us consider an example WPT system, as illustrated in Fig.~\ref{Fig:WPT_M(t)_circuit}. Time modulation of  mutual inductance can be realised by using nonlinear magnetic materials, mechanical movement, or with a digital control scheme. However, as we are interested in analysing the theoretical limits of the system performance, we do not focus on particular means of implementations. We consider an example WPT system with the following system parameters: $L_{1,2}=100~{\rm \mu H}$, $R_{1,2}=100~{\rm m \Omega}$, $\ws=2 \pi 100~{\rm krad/s}$, $C_{1,2}=1/(\ws^2L_{1,2})$, $\Zs=2~{\rm \Omega}$, and $\Rload=2~{\rm \Omega}$. The modulation frequency equals to the double of the source frequency $\ws$, i.e., $\wM=2\ws$, to ensure strong performance enhancement of the WPT system. First, we present efficiency $\eta$ and power ratio $\Rhload$ variations in a classical arrangement with a static mutual inductance (mutual inductance is not time-modulated). The results in Fig.~\ref{Fig:WPT_PTE_TPR_vscoupling} show the variation in transferred power ratio and efficiency of an unmodulated WPT system with respect to the coupling coefficient $k= {M_0}/\sqrt{L_1 L_2}$. We can observe that in conventional systems  efficiency increases with the increase of  coupling strength, however, the delivered power has a maximum at a particular coupling level ($k\approx0.033$ in this example), and it decreases at higher or lower coupling levels.  

\begin{figure}[!t]
	\centerline{\includegraphics[width= 0.75\columnwidth]{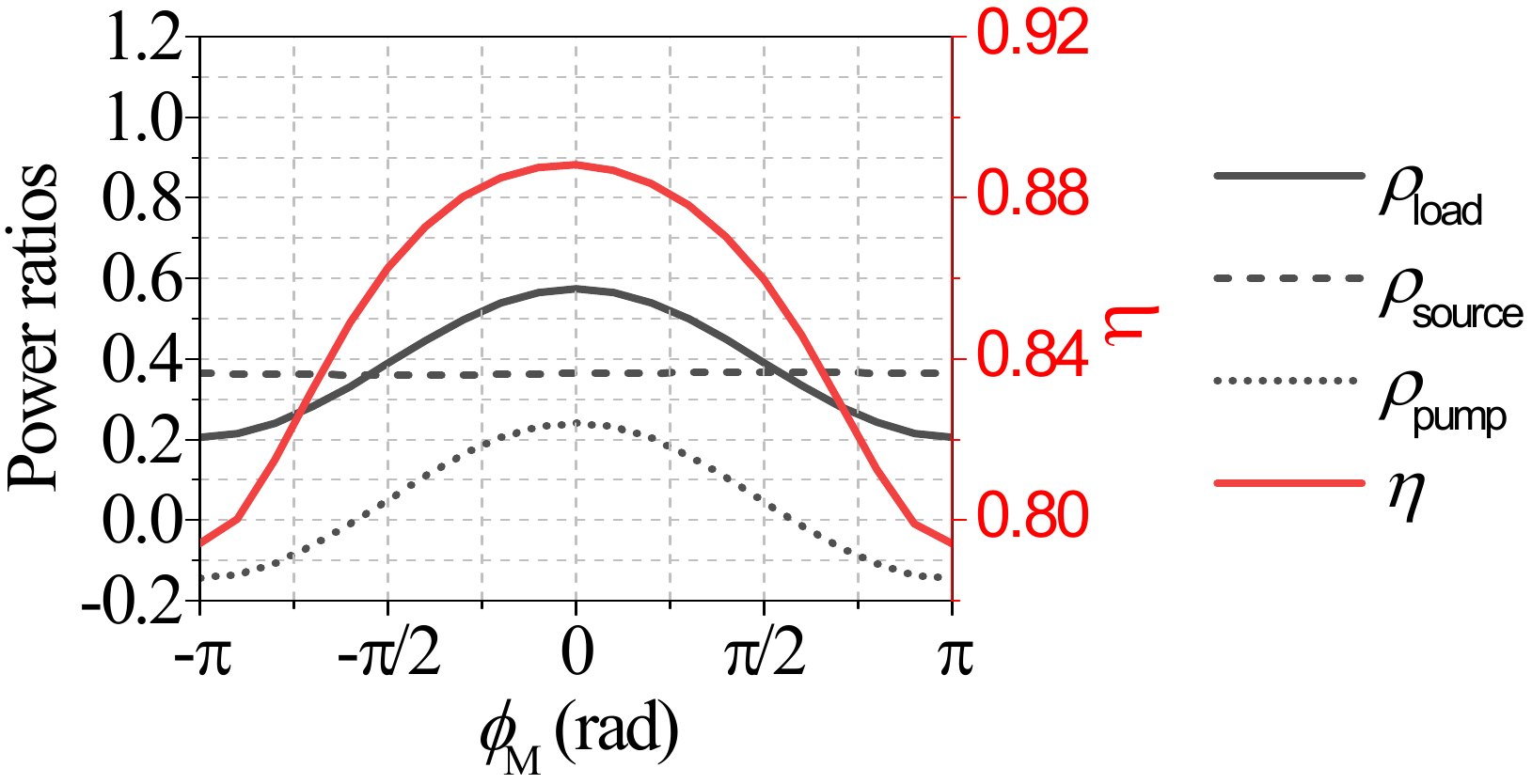}}
 		\caption{The variations of power ratios $\Rhload$, $\Rhsource$, $\Rhpump$ and transfer efficiency $\eta$ of the time-modulated WPT system with respect to the  modulation angle $\phiM$, with $m_{\rm M}=0.5$.} 
 		\label{Fig:WPT_PTE_TPR_vsPhase_M}
\end{figure}

Now, let us introduce  time modulation to the mutual inductance and analyse its effects on the transferred power ratio and efficiency. We choose $ {M_0}=0.1 \sqrt{ L_1 L_2}$, i.e., $k=0.1$, as a numerical example for the analysis.  First, the effect of the modulation phase $\phiM$ is analysed and the results are shown in Fig.~\ref{Fig:WPT_PTE_TPR_vsPhase_M}.  It should be highlighted that with time modulation of the mutual inductance, extra energy is pumped to the WPT system due to the action of the  time-modulation pump. Therefore, power ratios are calculated considering the power delivered to the load by two means: from the main source $\Vst$ and from the energy pump that modulates the mutual inductance. The source power ratio and pump power ratio are defined as $\Rhsource=\Psource / P_0$ and $\Rhpump=\Ppump / P_0$,respectively, where $\Psource$ is the power delivered from the main power source and $\Ppump$ is the power pumped from the time-modulation pump.

Importantly, the energy that comes from the pump due to the time modulation of the mutual inductance can be directly delivered to the receiver without incurring losses due to the internal source resistance or dissipation in the transmitter coil. Therefore, the transferred power to the load can be increased without reducing  efficiency due to the internal resistance of the source. This is the fundamental rationale behind the possibility to overcome the theoretical limit of  efficiency.

Depending on the modulation phase $\phiM$, the power flow from the pump can be either positive or negative, as seen from $\Rhpump$ in Fig.~\ref{Fig:WPT_PTE_TPR_vsPhase_M}. Positive pump power means that the energy is injected into the WPT system, while negative pump power means that energy is extracted from the main source to the pump.  Therefore, the modulation phase should be tuned properly.  It can be observed that the power ratio $\Rhload$ reaches its  maximum when the modulation of the mutual inductance is in-phase with the source voltage.  

\begin{figure}[!t]
	\centerline{\includegraphics[width=  0.75\columnwidth]{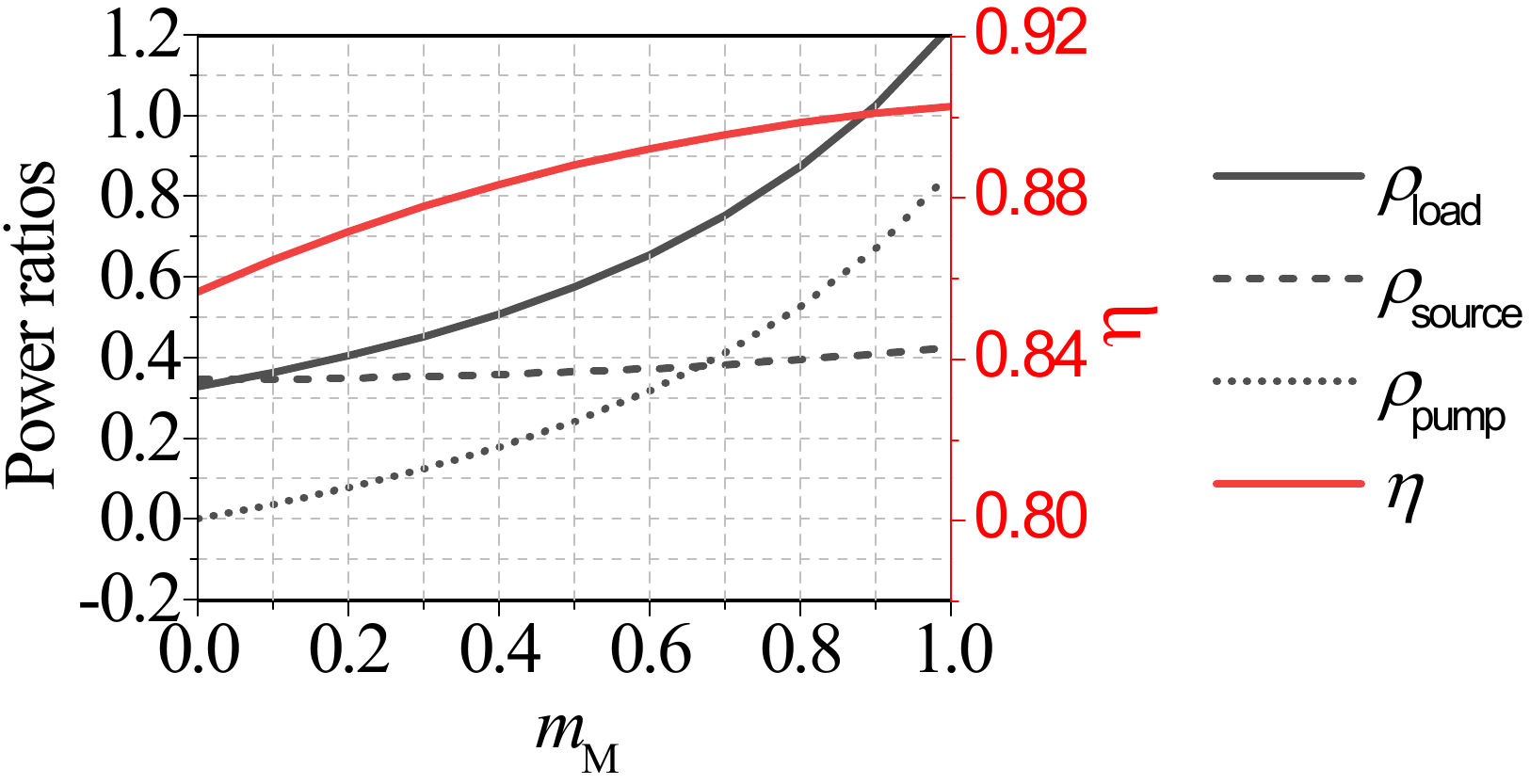}}
 		\caption{The variations of power ratios $\Rhload$, $\Rhsource$, $\Rhpump$, and transfer efficiency $\eta$ of the time-modulated WPT system with respect to modulation depth $m_{\rm M}$, with $\phiM=0$ and $M_0=0.1\sqrt{L_1L2}$.}
 		\label{Fig:WPT_PTE_TPR_vsModulation}
\end{figure}

The variations of power ratios and efficiency versus the  modulation depth is presented  in Fig.~\ref{Fig:WPT_PTE_TPR_vsModulation}. Interestingly, we can increase both efficiency and transferred power ratio by increasing the modulation depth. For example, when $m_{\rm M}=0.5$,  transferred power ratio increases from $32\%$ to $60\%$ while efficiency increases from $85\%$ to $89\%$ compared to the case without time modulation. This is not achievable by classical means of WPT optimizations such as impedance matching or frequency tuning because the losses inside the source will always be dominant under the  impedance-matched scenario. For instance, if we increase the coupling coefficient to $0.15$ in a static WPT system (which corresponds to the  maximum coupling in the  time-modulated WPT with $m_{\rm M}=0.5$), transferred power will be reduced to $16\%$ even if the efficiency can be increased to $91\%$.  We note that the performance will have a similar shape with a value shift when there is an offset of $M_0$ (change of the coupling coefficient $k$). We are able to overcome this theoretical limitation with  time modulation of the mutual inductance because we directly deliver additional energy to the receiver from the modulation pump. It should be highlighted that losses inside the pump that time-modulates the mutual inductance have not been taken into account in this analysis. These losses will inevitably reduce the additionally delivered power. However, in principle, the losses in modulation pump can be made very small, which is dependent on the particular realization method.  Importantly, bypassing the lossy circuits of the primary source and the parasitic resistance of the transmitting coil  allows us to overcome fundamental limitations on performance. We note that, in this particular example with $k=0.1$, the delivered power is mainly transferred via the fundamental harmonic, and the contribution from other harmonics are negligible. We expect that further explorations of the capabilities of time-modulated WPT such as the use of multiple-harmonic power transfer will open up completely new opportunities  for WPT developments.

\section{Time-modulated resistors for advanced antennas}
\label{sec:tmrapplic}

Next, we present an example of a  possible use of time-modulated resistors  for several potential applications, including broadband matching of antennas and enhancement of bandwidth of absorbers. For example, we can consider the situation where we modulate the internal resistance of a source that feeds a transmitting antenna, the radiation resistance of an antenna, or the load of a receiving antenna.   Although time varying resistors have been discussed in early literature on  modulators, it appears that their capabilities of generating virtual reactances for  broadband matching or absorption has not been identified before.
Knowing that time modulation of resistance can emulate reactance, we will explore this possibility to match an antenna in a wide frequency range. Recently, the use of time-modulated reactive elements was considered in paper \cite{TM-BandWidth}, where effective negative resistance was used for parametric amplification of the antenna current. The fundamental advantage of modulating resistance is that there is no need to introduce additional reactive elements that inevitably store reactive energy and increase the antenna quality factor. 
In this conceptual study we focus on fundamentally novel opportunities that open up due to time modulations of resistances, not going into details of specific implementations.  

\subsection{Effective impedance of the time-modulated resistor}

First, let us consider a time-modulated resistor $R(t)$ in an electrical circuit with resistance depending on time as  
\begin{equation}
    R(t)=R_0\Big(1 + \mM \cos(\wM t+\phiM)\Big),\label{eq:timevaryingR}
\end{equation}
where $R_0$ is the  nominal resistance, $\mM$, $\wM(=2\ws)$, and $\phiM$ are the modulation depth, modulation frequency and modulation phase respectively. If we assume that the coupling between higher-order harmonics is negligible, we can derive the effective impedance of the time-modulated resistor using the method presented in Section~\ref{sec:tmiprlc}. The first-order approximation result reads
\begin{eqnarray}
	R_{\rm R-eff}^{\omega_s} & \approx & R_0\left(1+\frac{1}{2} \mM \cos(2 \theta -\phiM)\right), \label{Eq:Leff_R(t)}   \\
	X_{\rm R-eff}^{\omega_s} & \approx & -\frac{1}{2} \mM R_0 \sin(2 \theta -\phiM).\label{Eq:Reff_X(t)}
 \end{eqnarray} 
From Eq.~\eqref{Eq:Reff_X(t)}, we can notice an interesting effect that the time-varying resistance can generate an effective reactance which shows a sinusoidal variation with respect to the modulation angle with the  amplitude of $\frac{1}{2}\mM R_0$. It appears that this effect has not been noticed before. It can be regarded as a dual phenomenon of the time-varying reactance (effective resistance can be generated by the modulation of capacitance or inductance, and its value is dependent on the modulation phase \cite{Gonorovsky}.
Under the assumption of negligible coupling between higher-order harmonics, it is possible to realize any reactance value within the range $\pm \frac{1}{2}\mM R_0$ by properly tuning the modulation angle. This is an interesting possibility, as it allows us to realize a tunable matching element without adding any additional reactive components. This is the reason for choosing the modulation frequency double of the signal frequency, which can provide this additional virtual reactance. Note that in contrast to reactances of any circuit formed by capacitors and inductors, this effective reactance does not depend on the frequency, offering a possibility to overcome the  fundamental limitations on the bandwidth of matching networks that follow from the Foster theorem.

\begin{figure}[!t]
\centerline{\includegraphics[width=\columnwidth]{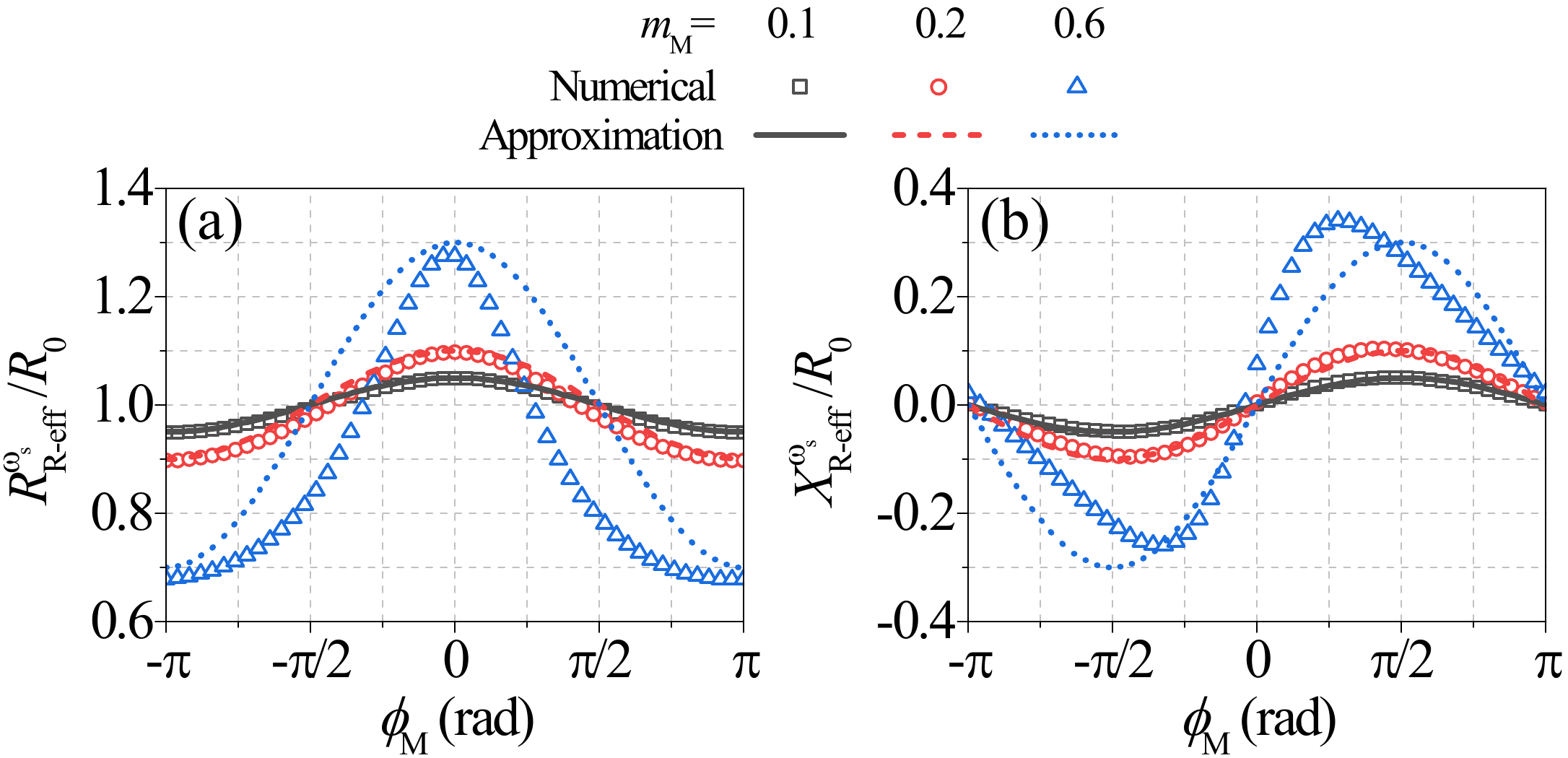}}
 	\caption{(a)~Effective resistance and (b)~effective reactance of time-modulated resistor versus the modulation phase $\phiM$.}
 	\label{Fig:R(t)_X_eff_vs_Phase}
\end{figure}

Let us now consider a numerical example of an $RLC$-circuit with a time-modulated resistor (same as that in Fig.~\ref{Fig:RL(t)C_circuit} but now the time-modulating element is the resistor) in order to further investigate the effects of coupling between higher-order modes. We use the following parameters for the numerical study: $L_0=100\,\,\mu$H, $R_0=100~\Omega$, $C_0=\frac{1}{\ws^2L_0}$, and $\ws=2 \pi 100$~krad/s. Variations of effective resistive and reactive impedances (normalized to $R_0$) with respect to the  modulation angle $\phiM$ are shown in Fig.~\ref{Fig:R(t)_X_eff_vs_Phase} for different modulation depths. We can see from  Fig.~\ref{Fig:R(t)_X_eff_vs_Phase} that the analytical first-order approximation agrees well with numerical results for small modulation depths ($\mM=0.01$ and $=0.1$) but deviates for higher modulation depths. This result is expected because the analytical approximation neglects higher-order harmonics that become strong at high modulation depth, and the phase angle of the current $\theta$ is not negligible for large $\mM$. However, these results show that even for an extremely high modulation depth, the minimum and maximum values of the  effective reactance do not deviate much from the analytical approximation. Therefore, one can still use time-modulated resistors to realize arbitrary reactances approximately within $\pm\frac{1}{2}\mM R_0$ by just modulating it at the proper phase angle $\phiM$.

\subsection{Broadband matching of antenna with time-modulated resistor}

Let us consider  a small loop antenna, represented by an $LR$-circuit with an inductance $L_0=1$~nH connected to a source with the internal  resistance of $100~\Omega$ ($R_0=100~\Omega$). Let us assume that we time-modulate the internal source resistance as in Eq.~\eqref{eq:timevaryingR}. According to the above discussion, we can fully compensate the reactance of the inductor $\omega L_0$ in a wide frequency range by using modulation with the correct phase angle. This means that the time-modulated resistor can be presented as an artificial negative reactance. Here, the time-modulation of the source resistance is  at the  double frequency of the excitation with the correct phase  angle such that the effective virtual reactance introduced by the modulated resistor is equal to $-\ws L_0$. This modulation can be easily automated with a simple feedback control system to ensure that the current through the inductor is in phase with the voltage across it.

\begin{figure}[!t] 
 \centerline{\includegraphics[width= 0.85\columnwidth]{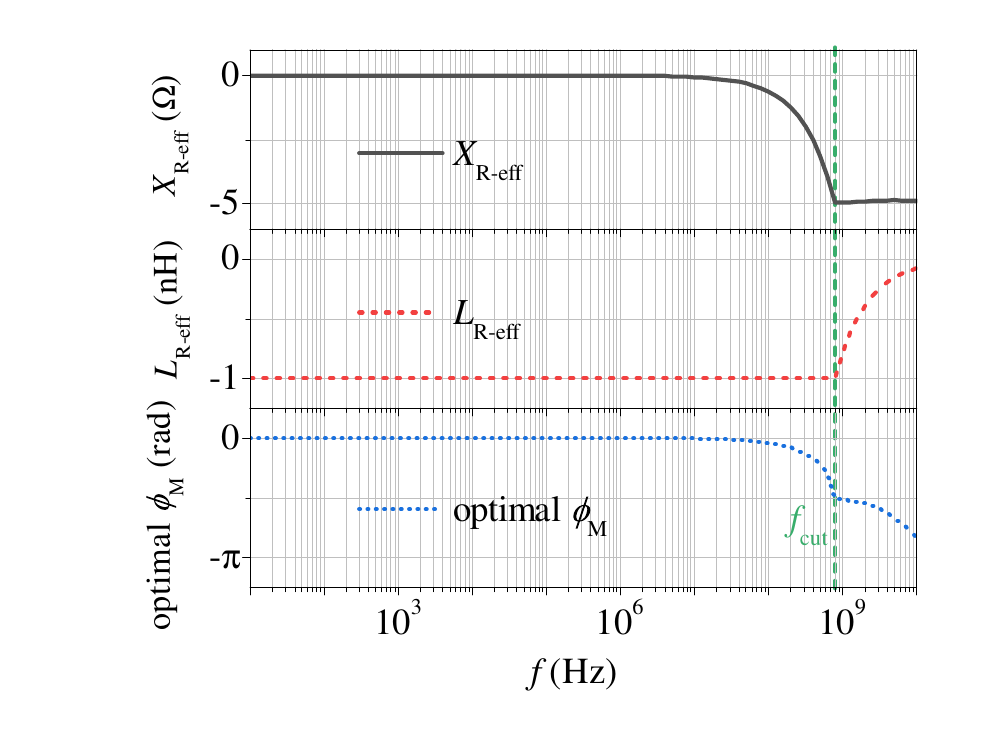}}
  	\caption{The effective reactance, inductance, and the optimal modulation phase of the time-modulated resistor with respect to the signal frequency, for achieving broadband matching of antenna($L_0=1$~nH, $R_0=100~\Omega$, $\mM=0.1$).}
  	\label{Fig:Negative_L_from_LR(t)}
	\vspace{-10pt}
\end{figure}

Example numerical results of realizing the broadband matching antenna with time-modulated resistance are shown in Fig.~\ref{Fig:Negative_L_from_LR(t)}, where the required optimal modulation phase is calculated to achieve impedance matching at different frequencies. The results show that  this non-Foster behavior can be achieved at frequencies up to around $800$~MHz.  The upper cut-off frequency of the range where we can fully compensate the loop inductance can be approximated as $f_{\rm  cut}=\frac{1}{2}\mM R_0/(2\pi L_0)$.
Similar considerations apply to small electric-dipole antennas, by using a $CR$-circuit.

 
However, the above analysis is only valid for a single-frequency source. When we want to use a time-modulated resistor to improve  communication antennas, which has a message signal on the carrier wave, we need to send multiple frequencies through the antenna, and the antenna bandwidth is an important performance indicator. Therefore, we need to study the situation when the source has multiple frequencies.  To this end, let us consider an amplitude-modulated signal 
 \begin{equation}
      \Vst=V_0 \cos(\ws t) (1+d \cos(\wmd t+\thetad)),
      \label{Eq.AmpModulatedFreq}
 \end{equation}
where $\ws=2\pi\fs$ is the carrier frequency, $\wmd=2\pi\fd$, $\thetad$ and $d$ are the angular frequency, phase and amplitude for the message signal. 
In this case, the source signal comprises three frequency components, $\ws$ and $\ws \pm \wmd$, and it is important to find the effective impedances at all three frequencies. 

\begin{figure}[!t] 
 \centerline{\includegraphics[width=  \columnwidth]{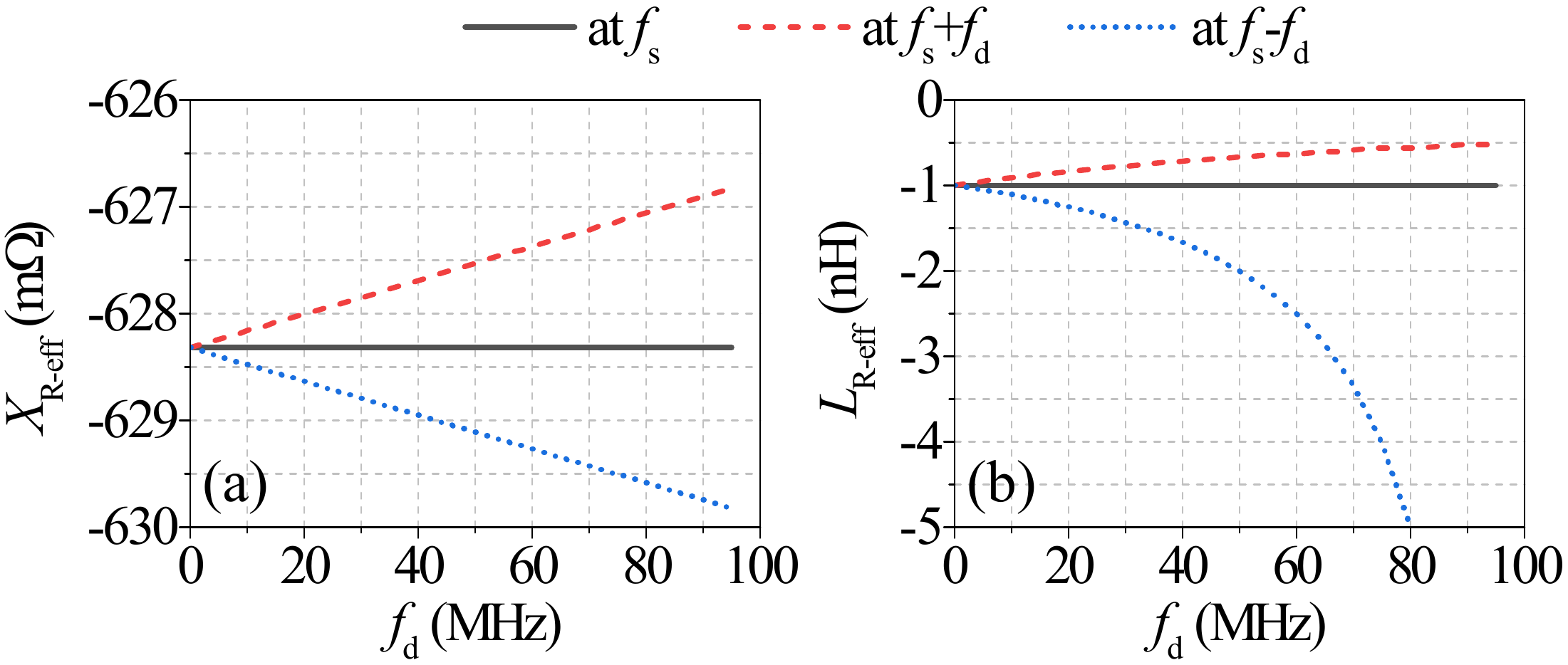}}
  	\caption{The effective reactance and inductance of the time-modulated resistor in Eq.~\eqref{eq:timevaryingR} at three different signal frequencies ($L_0=1$~nH, $R_0=100~\Omega$, $\mM=0.1$, $\fs=100$~MHz).}
  	\label{Fig:Negative_L_2freqSource_LR(t)1freq}
\end{figure}
\begin{figure}[!t] 
 \centerline{\includegraphics[width=  \columnwidth]{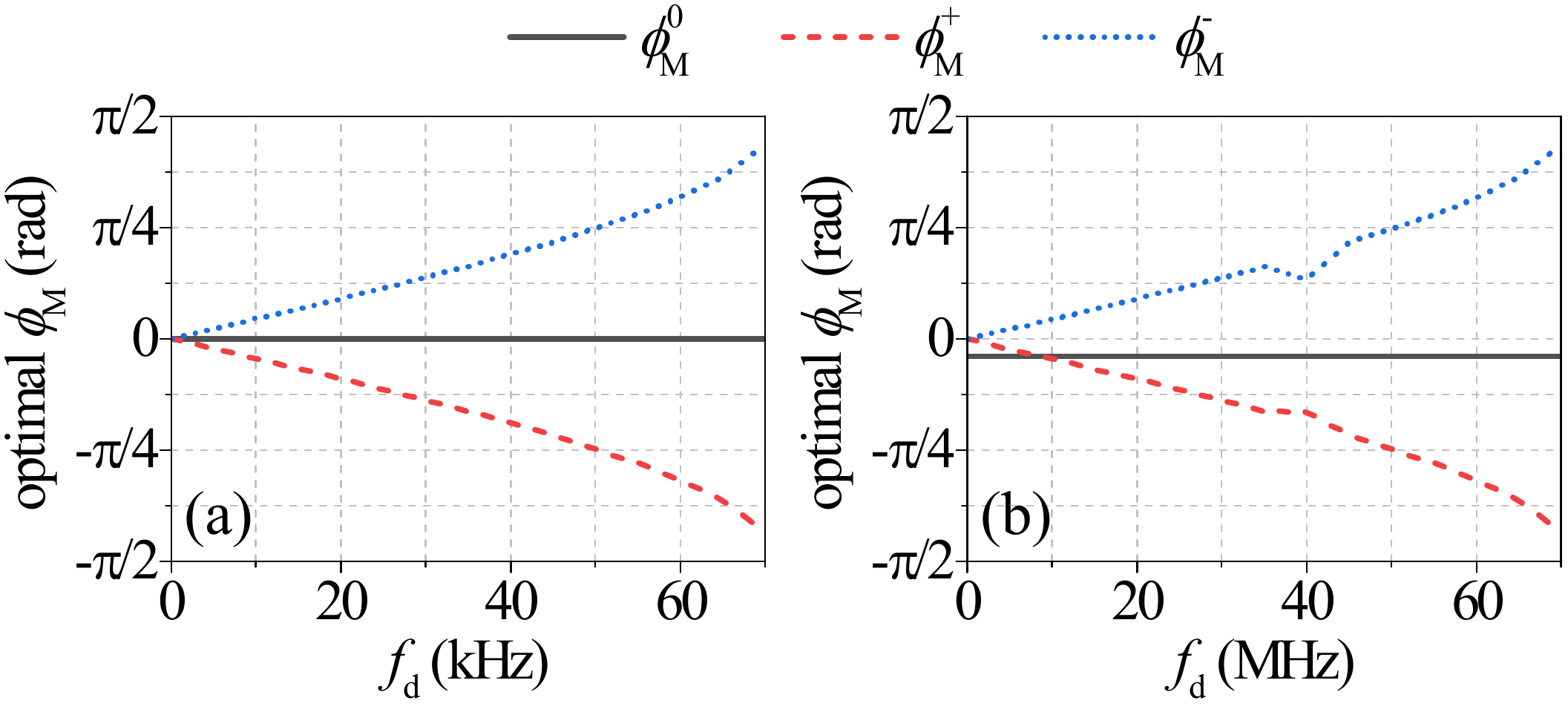}}
  	\caption{The optimal modulation phases realize perfect matching for communication antennas, with (a)~carrier frequency $\fs$ at $100$~kHz (with $\mM^0=0.1$ and $\mM^+=\mM^-=0.00001$), and (b)~carrier frequency $\fs$ at $100$~MHz (with $\mM^0=0.1$ and $\mM^+=\mM^-=0.01$).}
  	\label{Fig:Thriple_frequency_modulation_phase_vs_fd}
\end{figure}

Let us study the impedance characteristics of the time-modulated resistor introduced in the previous example (i.e., the time dependence of the modulated resistor is given by Eq.~\eqref{eq:timevaryingR}) that is excited by the considered amplitude-modulated signal $\Vst$ in Eq.~\eqref{Eq.AmpModulatedFreq} with $d=0.1$ and $\thetad=0$.  Figure~\ref{Fig:Negative_L_2freqSource_LR(t)1freq} shows the  effective reactance and inductance at  frequencies $\fs$, $\fs+\fd$, and $\fs-\fd$. We can see  that the effective 
reactance of the time-modulated resistor compensates the antenna reactance at the carrier frequency $f_s$; however, the antenna reactance at the other two frequencies (i.e., at  $\fs \pm \fd$) is not fully compensated. Therefore, in order to make the time-modulated antenna resonant at all the frequencies including the carrier frequency as well as the sideband frequencies, we introduce time-modulation of the antenna resistance at multiple frequencies, as
\begin{eqnarray}
R(t)=  R_0&\Big(1 + \mM^0 \cos(2 \ws t+\phiM^0) + \nonumber\\
      &  \mM^+ \cos(2 (\ws+\wmd) t+\phiM^+)+ \nonumber\\
      &  \mM^- \cos(2 (\ws-\wmd) t+\phiM^-)\Big),
\end{eqnarray}
\noindent where $\mM^{0,+,-}$ and $\phiM^{0,+,-}$ are the modulation depths and modulation phases at frequencies $2\ws,2(\ws \pm \wmd)$, respectively. That is,  we have  time modulation of the resistor at double of all the three signal frequencies. Basically, the resistance is time-modulated by an external force that is synchronized with the signal itself. It is interesting to note that using this modulation law, we can fully compensate the reactance of the antenna at all three frequencies of the signal spectrum.

Let us consider the same loop antenna example, modeled as $LR$-circuit as  studied above ($L_0=1$~nH and $R_0=100~{\rm \Omega}$), fed by an amplitude-modulated signal $\Vst$ as in Eq.~\eqref{Eq.AmpModulatedFreq}, with $d=0.1$ and $\thetad=0$. The results in Fig.~\ref{Fig:Thriple_frequency_modulation_phase_vs_fd} show the modulation phases required for perfect matching of the reactance at three frequencies $\fs$, $\fs+\fd$, and $\fs-\fd$, when we range the frequency $\fd$ of the message signal. The numerical calculations are presented for two cases: with $f_{\rm s}=100$~kHz and $f_{\rm s}=100$~MHz. It is clear from the results in Fig.~\ref{Fig:Thriple_frequency_modulation_phase_vs_fd} that the required effective negative inductance can be realized in a wide range of $f_{\rm d}$ values.    Since the system is linear, this approach is also valid for general (multi-tone) signals.


\section{Conclusion} 
\label{sec:conclusion}

This study of  time-varying electric elements (capacitor, inductor, and resistor) based on the introduced model of matrix circuit parameters has revealed  important features that appear due to temporal modulation of those elements. 
Developing these reasonably simple models for non-stationary elements is possible because these elements are linear and causal. Assuming that the components are electrically small at all relevant frequencies allows using bulk-component models, and if the signals and modulations are periodical functions of time, the models further simplify into matrix relations between vectors formed by the frequency harmonics of voltages and currents.  Importantly, the developed analytical models fully account for frequency dispersion of time-modulated components.

Applying these models to time-modulated mutual impedance between transmitting and receiving coils in inductive wireless power transfer systems, we have found that it is possible to improve transfer power ratio and power transfer efficiency beyond the known possibilities based on impedance matching in resonant scenarios. 

Considering the effects of time-modulated resistance, we have found a possibility to virtually match the load (for example, a small resonant antenna) to a source without the need to add any reactive matching networks or time-varying reactive components. This approach can be used also for bandwidth  enhancement of  receiving antennas and absorbers. 

To summarize, we would like to stress the key contributions of this work: 1. circuit-theory models and rigorous characterization of dispersive time-varying circuit elements; 2. introduction and analysis of temporally modulated resistance and mutual inductance; 3. revealing new possible applications of time-varying elements in wireless power transfer and antenna systems. We envision that these results may have significant impact on the future of this research direction.



\end{document}